\documentclass[5p,twocolumn,final]{elsarticle} 

\journal{Automatica}
\biboptions{numbers,sort&compress}

\makeatletter
\def\ps@pprintTitle{%
	\let\@oddhead\@empty
	\let\@evenhead\@empty
	\def\@oddfoot{}%
	\let\@evenfoot\@oddfoot
}
\makeatother

\usepackage{amssymb,amsmath,color,graphicx,epstopdf}
\usepackage[english]{babel}

\newproof{proof}{Proof}
 
\newtheorem{assumption}{Assumption}
\newtheorem{lemma}{Lemma}

\newtheorem{corollary}{Corollary}
\newtheorem{remark}{Remark}
\newtheorem{theorem}{Theorem}
\newtheorem{definition}{Definition} 

\newtheorem{problem}{Problem}

\newcommand{\R}{\mathbb R}

\newcommand{\N}{\mathbb N}

\newcommand{\Rplus}{\R_{\ge 0}}

\newcommand{\cA}{\mathcal A}

\newcommand{\cC}{\mathcal C}

\newcommand{\cE}{\mathcal E}

\newcommand{\cM}{\mathcal M}
\newcommand{\cN}{\mathcal N}
\newcommand{\cO}{\mathcal O}

\newcommand{\cR}{\mathcal R}
\newcommand{\cS}{\mathcal S}

\DeclareMathOperator{\setDistOp}{dist}

\newcommand{\x}{\times}
\newcommand{\st}{\,\mid\,}

\newcommand{\setdist}[2]{\setDistOp\left(#1,\;#2\right)} 
\newcommand{\und}{\underline}
\newcommand{\inv}{^{-1}}

\DeclareMathOperator{\dom}{dom}

\DeclareMathOperator*{\argmin}{argmin} 
\DeclareMathOperator*{\argmax}{argmax} 
\DeclareMathOperator{\card}{card}

\newcommand{\closure}[1]{\overline{#1}}

\newcommand{\sr}{^\star}

\allowdisplaybreaks
\newcommand{\projOp}[1]{\Pi_{#1}}
\newcommand{\proj}[2]{\projOp{#1}\left[#2\right]}

\newcommand{\e}{{\rm e}}

\newcommand{\M}{{\rm M}}
\newcommand{\uM}{\M\sr}
\newcommand{\lb}{\mu}
\newcommand{\ulb}{\und{\lb}}

 \newcommand{\EOP}{\ignorespaces~\hfill$\blacksquare$}

\begin{document}

\begin{frontmatter}

	\title{A Distributed Methodology for Approximate Uniform Global Minimum Sharing}

	\author[m] {Michelangelo Bin\corref{cor1}} 
	\author[m,t,e] {Thomas Parisini}
	
	\address[m]{Imperial College London, London, UK.}
	\address[t]{KIOS Research and Innovation Center of Excellence, University of Cyprus, Cyprus.}
	\address[e]{University of Trieste, Trieste, Italy.}
	
	\begin{abstract}
The paper deals with the distributed minimum sharing problem: a set of decision-makers compute the minimum of some local quantities of interest in a distributed and decentralized way by exchanging information through a communication network. We propose an adjustable approximate solution which enjoys several properties of crucial importance in applications. In particular, the proposed solution has good decentralization properties and it is scalable in that the number of local variables does not grow with the size or topology of the communication network. Moreover, a global and uniform (both in the initial time and in the initial conditions) asymptotic stability result is provided towards a steady state which can be made arbitrarily close to the sought minimum. Exact asymptotic convergence can be recovered at the price of losing uniformity with respect to the initial time.
	\end{abstract}

\end{frontmatter}


\section{Introduction}
\subsection{Problem Description, Objectives and Context}\label{sec.intro}

We consider the problem of computing the minimum of a set of numbers over a network, and we propose a distributed, iterative solution achieving \emph{global} and \emph{uniform}, albeit \emph{approximate}, asymptotic stability.  We are given a set $\cN$ of $N$ decision makers (or \emph{agents}), where each agent $i\in\cN$ is provided with a number $\M_i\in \R_{\ge 0}$ not known a priori by the others. The agents exchange information over a communication network with only a subset of other agents (called their \emph{neighborhood}).
The approximate minimum sharing problem  consists in the design of an algorithm guaranteeing that each agent asymptotically obtains a ``sufficiently good'' estimate of the quantity
\begin{equation}\label{d.uM}
\uM := \min_{i\in\cN} \M_i .
\end{equation}   
Clearly, ``$x_i=\uM,\ \forall i\in\cN$'' is also the {\em unique} solution to every  constrained optimization problem of the form
\begin{equation}\label{s.min_opt}
	\begin{aligned}
		&\max \,  \sum_{i\in\cN} \psi_i(x_i) \\ 
		&\quad x_i \le   \M_i , & \forall &i\in\cN\\
		&\quad x_i=x_j,& \forall & i,j\in\cN
	\end{aligned}
\end{equation}
obtained with $\psi_i$, $i\in\cN$, continuous and strictly increasing functions. Therefore, the minimum sharing problem is equivalent to the constrained distributed optimization problem \eqref{s.min_opt}, thus  intersecting  the wide research field of distributed optimization \cite{NotarstefanoTutorial}.

The problem of computing a minimum (or, equivalently, a maximum) over a network of decision makers is a classical problem in multi-agent control, with applications in distributed estimation and filtering, synchronization, leader election, and computation of network size and connectivity (see, e.g., \cite{Bullo2009,Santoro2006,nejad_maxconsensus_2009,iutzeler_analysis_2012,golfar_convergence_2019} and the references therein). 
Perhaps the most elementary existing algorithms solving the minimum sharing problem are   the \emph{FloodMax} \cite{Bullo2009} and the \emph{Max-Consensus} \cite{nejad_maxconsensus_2009,iutzeler_analysis_2012,golfar_convergence_2019}.  In its simplest form, Max-Consensus\footnote{For brevity, we only focus on  Max-Consensus. However,  the same conclusions applies also to the FloodMax.} requires   each agent $i\in\cN$ to store  an estimate $x_i\in\R$ of $\uM$  which is updated iteratively on the basis of the following update rule 
\begin{subequations}\label{s.ex.maxconsensus}
\begin{equation}\label{s.ex.maxconsensus_updatelaws}
	x_i^{t+1} =    \min_{j\in [i]} x_j^t ,\qquad  \forall i\in\cN ,
\end{equation}
with the initialization
\begin{equation}\label{s.ex.maxconsensus_initialization}
x_i^{t_0}= \M_i ,\qquad \forall i\in\cN,	
\end{equation}
\end{subequations}
where $t$ is the iteration variable, $t_0$ its initial value, and $[i]\subset\cN$ denotes  the neighborhood of agent $i$ (we assume $i\in[i]$). The update law \eqref{s.ex.maxconsensus_updatelaws} is decentralized and scalable, in that each agent needs only information coming from its neighbors and each agent stores only one variable. However, although   \eqref{s.ex.maxconsensus_updatelaws}  guarantees  convergence of each $x_i$ to $\uM$ when the estimates $x_i$ are initialized as specified in~\eqref{s.ex.maxconsensus_initialization}, \emph{convergence is not guaranteed for an arbitrary initialization}.  In fact,  if
\begin{equation}\label{s.ex.init2}
\exists i\in\cN \ {\rm s.t.}\ 	x_i^{t_0}<\uM, 
\end{equation}
 then the corresponding estimate $x_i^t$ produced by \eqref{s.ex.maxconsensus_updatelaws} satisfies $x^t_i<\uM$   all subsequent $t$, so that $x_i^t\to \uM$ cannot hold\footnote{In this specific case, we also observe that any  \emph{consensual} configuration (i.e., $x_i=x_j$ for all $i,j\in\cN$) is an equilibrium of \eqref{s.ex.maxconsensus_updatelaws}. This, in turn, is intimately linked to the unfeasibility result of \cite[Theorem 3.1.1]{Santoro2006}, and to the \emph{detectability} issues  appearing in many control problems, such as   \emph{Extremum Seeking}  \cite{Ariyur2003,Tan2006}.}. Therefore, since convergence to $\uM$ holds only for some specific initial values $x_i^{t_0}$, the Max-Consensus algorithm~\eqref{s.ex.maxconsensus} is not  \emph{globally convergent}.
While there are application domains for which attaining global convergence is not strictly necessary, there are many others in which it is a crucial requirement.  This is the case, for instance, when the quantities $\M_i$ can  change at run time (see  the two use-cases illustrated in Section~\ref{sec.app}). To see how this may be a problem for the update law  \eqref{s.ex.maxconsensus}, assume by way of example that the estimates $x_i^t$ have reached at a given $t_1$ the value $\uM$, i.e. $x_i^{t_1}=\uM$ for all $i\in\cN$, and assume that there is a unique $k\in\cN$ such that $\uM=\M_k$. Now, suppose that at some $t_2>t_1$ the value of $\M_k$ increases, thus determining an increment also of $\uM$. Then, the condition \eqref{s.ex.init2} holds for $t_0=t_2$ so as, in view of the discussion above, the update law \eqref{s.ex.maxconsensus_updatelaws} fails to track the new minimum. 

Global attractiveness is not the only desirable property one may be interested in when the minimum sharing problem is considered over large networks with possibly changing conditions. In fact, a crucial role is also played by
\begin{enumerate} 
	\item \emph{Uniformity of the convergence}: the convergence rate does not depend on the initial value $t_0$ of the iteration variable and is constant over compact subsets of initial conditions.
	\item  \emph{Stability of the steady state}: ensures that small variations in the parameters and initial conditions map into small deviations from the unperturbed trajectories.
	\item  \emph{Scalability}:  the number of variables stored by each agent does not grow with the network size or the number of interconnections.
	\item \emph{Decentralization of the updates}: the update law of each agent uses only local information  and depends on parameters that are     independent   from those of the other agents.
\end{enumerate}	 
Indeed,  uniform global attractiveness and stability of the steady  state confer  robustness against uncertain and time-varying conditions and parameters (see  e.g. \cite[Chapter~7]{Goebel2012}), making the minimum sharing method suitable for  applications in which the quantities $\M_i$ vary in time. Moreover, scalability and decentralization  enable the application to large-scale networks. In this direction, in this paper we look for a novel solution to the minimum sharing problem having  scalability and decentralization properties similar to those of Max-Consensus~\eqref{s.ex.maxconsensus}, but, in addition, possessing the aforementioned globality, uniformity and stability properties.

\subsection{Motivating Applications}\label{sec.app}

Our methodology is   motivated by two application contexts described below. In both cases, a key element consists in solving an instance of the minimum-sharing problem \eqref{s.min_opt} in which the parameters~$\M_i$, hence the minimum $\uM$, may change over time. In this contexts, (i) global attractiveness allows to track the changing minimum $\uM$, (ii)~uniformity of convergence guarantees that the convergence rate is always the same  and  does not decrease with time, and   (iii) stability guarantees that relatively small variations of the parameters lead to small transitory deviations from the optimal steady state.

\subsubsection{Cooperative Control of Traffic Networks}
Consider a traffic network consisting of a set of vehicles driving on a highway  in an intense traffic situation. 
Some of the vehicles have self-driving capabilities, and we can assign their driving policies.
The other vehicles are instead human-driven and, thus, they are not controlled. The whole traffic network is seen as a \emph{plant} that, when not properly controlled, may exhibit undesired behaviors, such as   ghost jams. The control goal consists in finding a control policy, distributed among the self-driving vehicles, which guarantees that the ``closed-loop'' traffic network  behaves properly, leading to a smooth traffic flow where all the vehicles hold a common maximal cruise speed. At each time, the maximum attainable cruise speed of each vehicle $i$ is constrained by a \emph{personal maximum value},   denoted by $\M_i$,   which may  depend on mechanical constraints, on
the traffic conditions, on standing speed limitations,  or  other exogenous factors. 
A key part of the control task consists in the distributed computation of the   maximum common cruise speed,   $\uM$, compatible with all the personal   velocity constraints. 
At each time, the problem of estimating $\uM$ is  an instance of~\eqref{s.min_opt}, whose solution is precisely~\eqref{d.uM}. 

\subsubsection{Dynamic Leader Election}
Another important motivating application is the distributed \emph{leader election} problem in dynamic networks, which shares many similarities with the previous application. Single-leader election has been proved to be an unsolvable problem in general, even under bi-directionality, connectivity, and total reliability assumptions on the communication networks \cite[Theorem 3.1.1]{Santoro2006}. A standard additional assumption  making the problem well-posed  is that each agent is characterized by a \emph{unique identifier}~$\M_i$. Hence, the problem of leader election can be cast as finding  the minimum, $\uM$, of such identifiers. The agent whose identifier coincides with $\uM$ declares itself the leader, the others the followers.

\subsection{Related Works and State of the Art}\label{sec.literature}
Classical algorithmic approaches to the minimum sharing  problem  in arbitrary networks have been developed in the context of distributed algorithms and robotic applications. They  include  the \emph{FloodMax} \cite{Bullo2009}, the \emph{Max-Consensus}~\cite{nejad_maxconsensus_2009,iutzeler_analysis_2012,golfar_convergence_2019} (see \eqref{s.ex.maxconsensus}), the \emph{MegaMerger} \cite{Gallager1983}, and the \emph{Yo-Yo} algorithm. See \cite{Santoro2006,Bullo2009} for a more detailed overview. Some of these approaches, such as the basic   Max-Consensus~\eqref{s.ex.maxconsensus}, have nice  scalability and decentralization properties: the update laws do not depend on \emph{centralized quantities}, such as parameters  that need to be  known in advance by all the agents, and employ  a number of local variables which does not grow with the network size or topology. However, all such approaches require a correct initialization or a pre-processing synchronization phase, which are   undesired limitations  in applications of interest such as, for example, the ones discussed in Section~\ref{sec.app}. 

If the minimum sharing problem is cast in terms of the optimization problem \eqref{s.min_opt}, then one can rely on a well-developed literature on discrete-time distributed optimization  (see \cite{NotarstefanoTutorial} for a recent overview). If the functions $\psi_i$ in~\eqref{s.min_opt} are convex, indeed,  different approaches can be used, such as {consensus-based (sub)gradient methods}  \cite{nedic_distributed_2009,nedic_constrained_2010,lobel_distributed_2011,shi_extra_2015,shi_proximal_2015,yuan_convergence_2016},  {second-order methods} \cite{varagnolo_newton-raphson_2016,mokhtari_network_2017}, projected \cite{xie_distributed_2018} and primal-dual \cite{zhu_distributed_2012,chang_distributed_2014} methods with inequality constraints, methods based on the distributed Alternate Direction Method of Multipliers (ADMM) \cite{Boyd2011,mota_d-admm_2013,shi_linear_2014,jakovetic_linear_2015,ling_dlm_2015,chang_proximal_2016,makhdoumi_convergence_2017,NotarstefanoTutorial,bastianello_asynchronous_2020}, and methods based on gradient tracking \cite{xu_augmented_2015,nedic_achieving_2017,nedic_geometrically_2017,qu_harnessing_2018,xi_add-opt_2018,Bin2019}.
Gradient methods typically achieve global attractiveness. However, among the cited references only \cite{nedic_constrained_2010} deals with constrained problems with different \emph{local constraints}\footnote{By the term ``local constraints'' we refer to private constraints an agent may have on its own variables that do not depend on the other agents' variables, e.g. the constraints $x_i\le \M_i$ in \eqref{s.min_opt}.} such as \eqref{s.min_opt}. Yet,  \cite{nedic_constrained_2010}  requires a vanishing stepsize, which makes convergence not uniform. Gradient methods employing a fixed stepsize  thus guaranteeing uniformity  are given in  \cite{nedic_distributed_2009,lobel_distributed_2011,shi_extra_2015,shi_proximal_2015,yuan_convergence_2016,varagnolo_newton-raphson_2016,mokhtari_network_2017}. However, they do not cover constrained problems of the kind~\eqref{s.min_opt}. Moreover, the first-order methods in  \cite{nedic_distributed_2009,lobel_distributed_2011,yuan_convergence_2016}  lead to an approximate  convergence result in which the convergence speed and the approximation error need to be traded off. This, in turn, is consistent with our results  in which a compromise is  more generally  established between uniformity, approximation error and convergence rate. The approaches~\cite{xie_distributed_2018,zhu_distributed_2012,chang_distributed_2014} deal with inequality constraints including Problem~\eqref{s.min_opt}. Nevertheless, they require a correct initialization and, hence, they do not provide global attractiveness. The same issue applies to gradient-tracking methods~\cite{xu_augmented_2015,nedic_achieving_2017,nedic_geometrically_2017,qu_harnessing_2018,xi_add-opt_2018,Bin2019} (which, anyway, are developed for unconstrained problems), and also for the ``node-based'' formulations of ADMM   \cite{mota_d-admm_2013,shi_linear_2014,jakovetic_linear_2015,makhdoumi_convergence_2017,ling_dlm_2015}. Instead, the ``edge-based'' formulations of ADMM (e.g. \cite[Section 3.3]{NotarstefanoTutorial}, \cite{bastianello_asynchronous_2020}) do not suffer from this initialization issue, and they provide a solution which is global and uniform. Nevertheless, the number of variables that each agent has to store grows with the dimension of its neighborhood, thus incurring in  scalability issues. 
Moreover,  
stability is not usually considered in the analysis of the aforementioned designs,
and typically   the update laws employ   coefficients (e.g. stepsizes) which must be common\footnote{Exceptions  are given in the gradient-tracking designs of \cite{xu_augmented_2015,nedic_geometrically_2017}, where  agents employ uncoordinated stepsizes. In both the designs, the discrepancy between the stepsizes must be small enough. Hence, these results may be seen as a  ``robustness'' property relative to variations of the stepsizes with respect to their average. In turn, this property  comes \emph{for free}  if the algorithm is proved to be \emph{asymptotically stable} with a common stepsize (see, e.g., \cite[Chapter 7]{Goebel2012}).} to all agent (i.e., they are \emph{centralized} quantities).

\subsection{Contributions \& Organization of the Paper} \label{sec.contribution}
We propose a new approach to the minimum sharing problem that provides an adjustable \emph{approximate} (or \emph{sub-optimal} in terms of \eqref{s.min_opt}) solution enjoying   the globality, uniformity, scalability and decentralization properties stated in Section~\ref{sec.intro}, which  do not seem to be possessed altogether by any   existing algorithm.   The proposed update laws have the form
\begin{equation}\label{s.xi_fi}
x_i^{t+1} = f_i(t,x^t),
\end{equation}
for some suitable functions $f_i$, where $x_i\in\R$ represents the estimate of $\uM$ stored by agent $i$, and $x:=(x_i)_{i\in\cN}$ is the aggregate estimate. As formally specified later on in Section~\ref{sec.comm}, the actual structure of the functions $f_i$ encodes the decentralization constraints, allowing an agent update to depend only on the estimates of a subset of other agents (see   Remark~\ref{rmk.decentralization}).
We show that all the estimates $x_i$ converge, globally and uniformly, to a stable neighborhood of $\uM$ whose size can be reduced arbitrarily around $\uM$  by suitably tuning some control parameters. 
More precisely, the proposed approach enjoys the following  properties:
\begin{enumerate}[(a)]
	\item The algorithm is  distributed and scalable, since the only one variable is stored for each agent.
	\item\label{item.dec} The update law of each agent employs a gain which can be tuned independently from the others.
	\item The    estimates $x_i$  converge globally and uniformly to a  stable  steady state which    can be made arbitrarily close to  $\uM$.
	\item Exact convergence (i.e., all the  estimates  converge to $\uM$) can be achieved, at the price, however, of losing uniformity. 
\end{enumerate}
In view of Item  \eqref{item.dec}, the proposed method has good decentralization properties compared to most of the approaches mentioned in Section~\ref{sec.literature}. Nevertheless, we underline that the proposed method is not fully decentralized, as the agents are supposed to know a lower-bound on $\uM$ (Assumption~\ref{ass.M_eps}) which explicitly enters in the update laws.

The paper is organized as follows. After providing preliminary   definitions  and remarks in Section~\ref{sec.prelim}, in Section~\ref{sec:min:share} we formulate the minimum-sharing problem  and we describe the proposed solution methodology. The main convergence results are given in Section~\ref{sec:conv} and proved in Section~\ref{sec.proof}. Finally, numerical results and concluding remarks   are reported in Sections~\ref{sec:simul} and~\ref{sec:concl}, respectively.


\section{Preliminaries} \label{sec.prelim}
\subsection{Notation} We denote by $\R$ and $\N$ the set of real and natural numbers respectively. If $a\in\R$, $\R_{\ge a}$ denotes the set of all real numbers larger or equal to $a$, and similar definitions apply to other ordered sets and ordering relations. We denote by $\card A$ the cardinality of a set $A$. If $A,B\subset \R$, $A\setminus B:=\{ a\in A\st a\notin B \}$ denotes the set difference between $A$ and $B$. We identify singletons with their unique element and, for a $b\in\R$, we thus write $A\setminus b$ in place of $A\setminus \{b\}$. We denote norms by $|\cdot|$ whenever they are clear from the context. With  $A\subset\R^n$  and   $x\in\R^n$,  $\setdist{x}{A}:= \inf_{a\in A}|x-a|$ denotes the distance from $x$ to $A$. Sequences indexed by a set $S$ are denoted by $(x_s)_{s\in S}$.
For a non-empty interval $[a,b]\subset \R$, we define the projection map $\projOp{[a,b]}:\R\to[a,b]$   as $\projOp{[a,b]}(s) := \min\{\max\{ s,\, a  \},\, b\}$.  A function $f:\R^n\to\R^m$, $n,m\in\N$, is \emph{locally bounded} if $f(K)$ is bounded for each compact set $K\subset\R^n$. 
In this paper, we consider discrete-time systems  whose solutions are signals defined on a non-empty subset $\dom x$ of $\N$. For ease of notation, we will use $x^t$ in place of $x(t)$ to denote the values of a signal $x$. With $t_0\in\N$, we say that $x$ \emph{starts at $t_0$} if $\min \dom x = t_0$.

\subsection{Communication Networks}\label{sec.comm}
Throughout the paper, $\cN$   denotes the (finite) set of agents in the network, and we let $N:=\card\cN$. 
The network communication constraints are formally captured by the   concept of ``communication structure''  defined below\footnote{A common way to define a communication structure on $\cN$ is to consider an undirected graph $(\cN,\cE)$ with vertices set equal to  $\cN$ and  edges set $\cE\subset\cN\x\cN$ such that if $(i,j)\in\cE$ then agents $i$ and $j$ can communicate. In this case,  $[i]:=\{i\}\cup\{ j\in\cN\st (j,i)\in\cE\}$.}.

\begin{definition}\label{def.com_struct}
	A \emph{communication structure} on $\cN$ is a sequence $\cC=([i])_{i\in\cN}$  of subsets $[i]$ of $\cN$ which satisfy $i\in[i]$.
\end{definition}
For each $i\in\cN$, the set $[i]$  is called the \emph{neighborhood} of $i$. A \emph{communication  network}  is a pair $(\cN,\cC)$, in which $\cN$ is a set and  $\cC$ is a communication structure on $\cN$.

For a given $I\subset\cN$, we define the sequence of sets
\begin{equation}\label{d.nbds}
\begin{array}{lcl}
[I]^0 &:=& I \\{} 
[I]^n &:=&  \bigcup_{j\in [I]^{n-1}} [j],\quad  n\in\N_{\ge 1}
\end{array}
\end{equation}
so as, in particular, $[\{i\}]^1=[i]$. If $I=\{i\}$ is a singleton, we use the short notation $[\{i\}]^n=[i]^n$. 
Moreover,  for $n,m\in\N$  we let
\begin{equation*}
[I]_m^n := [I]^n \setminus [I]^m.
\end{equation*}

We consider networks that are \emph{connected} according to the following definition.
\begin{definition}\label{def.Iconnected}
	With $I\subset\cN$, a communication network $(\cN,\cC)$ is said to be   $I$-\emph{connected} if there exists $n_I\le N$ such that $[I]^{n_I} = \cN$. 
\end{definition}
The notion of $I$-connectedness is in general weaker than   usual  \emph{strong connectedness}, which requires the existence of  a path between any two agents.  
Later on, we shall assume    that $\cN$ is given a communication structure $\cC$  which is $I\sr$-connected for a specific subset $I\sr\subset\cN$. For the   purpose of   analysis, this communication structure is assumed static. Likewise also the quantities $\M_i$ are supposed constant. In fact, this corresponds to a well-defined ``nominal setting'' for the proposed method in which we can prove the desired uniform global attractiveness and stability properties. Proving such   properties in the nominal case, in turn,  guarantees that the proposed method can be applied also to relevant classes of problems where the communication structure and the parameters $\M_i$ (hence, their minimum $\uM$) may change over time. Indeed, as already mentioned in Section~\ref{sec.intro}, uniform global attractiveness and stability ensure a proper approximate tracking of a time-varying minimum $\uM$ provided that its dynamics is sufficiently slow. 	Moreover, classical results in the context of control under different time-scales  (see,  e.g., \cite{Kokotovic1999,Teel2003,Tan2006,Wang2012}) also guarantee good tracking performances under changes of the communication structure $\cC$   that are, on average, sufficiently slow with respect to the dynamics of the update laws.  
In this respect, Section~\ref{sec:simul} provides numerical results in a scenario in which the communication structure and the numbers $\M_i$ are subject to  impulsive changes separated by relatively large  intervals of time.

\subsection{Stability and Convergence    Notions}\label{sec.convergence}
We consider discrete-time systems of the form
\begin{equation} \label{pre.s.x}
x^{t+1} = f(t,x^t),
\end{equation}
with state $x^t\in\R^n$, $n\in\N$.
Given a  closed set $\cA\subset\R^n$, we say that $\cA$ is \emph{stable} for \eqref{pre.s.x} if for each $\epsilon>0$ there exists $\delta(\epsilon)>0$ such that every solution of~\eqref{pre.s.x} satisfying $\setdist{x^{t_0}}{\cA}\le \delta(\epsilon)$ also satisfies $\setdist{x^{t}}{\cA}\le \epsilon$, for all $t\ge t_0$.
We say that   $\cA$ is \emph{attractive} for \eqref{pre.s.x} if there exists an open superset $\cO$ of $\cA$ and, for every $t_0\in\N$, every solution $x$ to \eqref{pre.s.x} with $x^{t_0}\in \cO$, and every $\epsilon>0$, there exists $t\sr(t_0,x^{t_0},\epsilon)\in\N$, such that $\setdist{x^t}{\cA}\le \epsilon$ holds for all $t\ge t_0+t\sr(t_0,x^{t_0},\epsilon)$.
Different qualifiers can enrich this attractiveness property. In particular, the set $\cA$ is said to be:
\begin{itemize}
	\item  \emph{Globally attractive} if $\cO=\R^n$.
	\item \emph{Finite-time attractive} if the condition ``$\epsilon>0$'' can be replaced by ``$\epsilon\ge 0$''.
	\item \emph{Uniformly attractive in the initial time} $t_0$ if   the map $t\sr(\cdot)$ does not depend on $t_0$. 
	\item \emph{Uniformly attractive in the initial conditions} $x^{t_0}$ if for each $(t_0,\epsilon)\in\N\x\R_{\ge 0}$,  the map $t\sr(t_0,\cdot,\epsilon)$ is locally bounded.
	
	\item \emph{Uniformly attractive} if it is both uniformly attractive in the initial time and in the initial conditions.
	
	\item  {\emph{$\epsilon$-approximately attractive}} (with $\epsilon>0$)  if the set $\{ x\in\R^n\st \setdist{x}{\cA}\le \epsilon   \}$ is attractive.
 
\end{itemize} 
If $\cA$ is both stable and attractive, it is said to be \emph{asymptotically stable}. 
Moreover,   with $(f_\gamma)_{\gamma\in \Gamma}$ representing a family of functions $f_\gamma:\N\x\R^n\to\R^n$ indexed by a set $\Gamma$, consider the family of systems
\begin{equation}\label{pre.s.xa}
 x^{t+1} = f_\gamma(t,x^t),\qquad\gamma\in \Gamma.
\end{equation}
Then, we say that the set $\cA$ is \emph{practically attractive} for the family \eqref{pre.s.xa}, if for each $\epsilon>0$, there exists $\gamma\sr(\epsilon)\in\Gamma$ such that the set $\cA$ is $\epsilon$-approximately attractive for the system~\eqref{pre.s.xa} obtained with $\gamma=\gamma\sr(\epsilon)$.

\section{Distributed Minimum Sharing}\label{sec:min:share}

 \subsection{Problem Formulation}
 
We are given a  communication network $(\cN,\cC)$. Each agent $i\in\cN$ is provided with a number $\M_i$, not known a priori by the others, and it stores and updates a local estimate $x_i\in\R$ of the quantity $\uM$ defined in \eqref{d.uM}. Thus, the problem at hand consists in designing an update law for each agent $i\in\cN$ of the form \eqref{s.xi_fi}
 such that the resulting estimates $x_i^t$ converge to $\uM$, in some of the senses defined in Section \ref{sec.convergence}. The resulting family $f:=(f_i)_{i\in\cN}$ is called the \emph{distributed methodology}. In the following,  we let $x:=(x_i)_{i\in\cN}$ and we compactly rewrite \eqref{s.xi_fi} as
\begin{equation}\label{s.x}
x^{t+1} = f(t,x^t).
\end{equation}
As each agent is allowed to exchange information only with the  agents belonging to its neighborhood $[i]$, the functions $f_i$ must respect this constraint. This is formally expressed by the following definitions.
\begin{definition}
With $V\subset\cN$, a function $g$ on $\N\x\R^N$ is said to be \emph{adapted to $V$} if  it satisfies $g(t,x)=g(t,z)$  for every $t\in\N$, and every $x,z\in\R^N$ satisfying $x_i=z_i$ for all $i\in V$.
\end{definition}
\begin{definition}\label{def.decentralized}
The function $f=(f_i)_{i\in\cN}$ is said to be $\cC$-\emph{decentralized} if, for each $i\in\cN$, the map $f_i$ is adapted to~$[i]$.
\end{definition}
Then, the  \emph{distributed minimum sharing} problem  is defined as follows.
\begin{problem}\label{prob.1}
Design a $\cC$-decentralized function  $f$, such that the set
\begin{equation}\label{d.A}
\cA := \{\uM\}^N
\end{equation}
 is   globally attractive for \eqref{s.x}.
\end{problem}

\begin{remark}\label{rmk.decentralization}
	We stress that, if $f$ is $\cC$-decentralized, then each function $f_i$ in \eqref{s.xi_fi} depends only on $(x_j)_{j\in [i]}$ and not on the whole state $x$. 
\end{remark}

\begin{remark}
Depending on the additional qualifiers that may characterize the attractiveness property of $\cA$ in Problem \ref{prob.1}, we may have solutions to Problem \ref{prob.1}   in ``different senses''. 
In the forthcoming section, we propose a methodology obtaining both global attractiveness   and global  uniform  {practical} attractiveness of $\cA$, depending on the value of some user-decided control parameters. We will  show that a compromise between how close we can get to $\cA$ and uniformity in the initial time is necessary. In particular, we show that  attractiveness is possible only at the price of losing uniformity in the initial time, and that, if such property is needed, then global  practical  uniform attractiveness is the best we can achieve. 
\end{remark}

\subsection{Standing Assumptions}
We consider Problem \ref{prob.1} under two  main assumptions specified hereafter. We define the set
\begin{equation}\label{d.Isr}
\begin{array}{lcl}
I\sr &:=& \displaystyle\argmin_{i\in\cN} \M_i .
\end{array}
\end{equation} 
With the following assumption, we require the communication network to be connected with respect to  $I\sr$.

\begin{assumption}[Connectedness]\label{ass.connected}
	The communication network $(\cN,\cC)$ is $I\sr$-connected in the sense of Definition~\ref{def.Iconnected}. 
\end{assumption}

The second assumption, instead, requires   each agent to know a lower-bound on $\uM$.
\begin{assumption}[Consistency]\label{ass.M_eps}
	Each agent $i\in\cN$ knows a number $\lb_i\in\R_{>0}$ such that $\lb_i\le\uM$.
\end{assumption}
It is worth noting that Assumption \ref{ass.M_eps} is a ``centralized'' assumption, in that it asks each agent to know a lower bound on the common, unknown quantity $\uM$. Nevertheless, it introduces almost no loss of generality   in different applications of interest, including those mentioned in Section \ref{sec.app},  where knowing a lower-bound on $\uM$ is a mild requirement. For instance, in both the traffic control and leader election problems we can assume that the quantities $\M_i$ are integers, so that ``$\lb_i\in(0,1)$ for all $i\in\cN$'' is a feasible choice  requiring no further knowledge on $\uM$.
 Furthermore, this assumption is  not in principle needed if an approximate or practical attractiveness result is sought. In fact,  if for some $I\subset \cN$, $\epsilon:=\max_{i\in I}\lb_{i}>\uM$, then $\uM\in[0,\epsilon)$, and, as clarified later on by the asymptotic analysis, we are able to  claim that the set $[0,\epsilon]^N$ (which includes $\uM$) is practically attractive for $x$, with $\epsilon$, however, that  can be made arbitrarily small by choosing $\lb_i$ accordingly. 
 
 In   the following we let
 \begin{equation}\label{d.ulb}
 \ulb  :=  \displaystyle\min_{i\in\cN} \lb_i.
 \end{equation}

\subsection{The Update Laws}

The proposed update law is obtained by choosing $f$ so that, for   each $i\in\cN$,  Equation  \eqref{s.xi_fi}  reads as follows\footnote{Recall that $\projOp{[a,b]}(s) := \min\{\max\{ s,\, a  \},\, b\}$.}
\begin{equation}\label{s.xi}
x_i^+ = \proj{[\lb_i,\, \M_i]}{\e^{h_i^t} x_i + k_i \sum_{j\in[i]}\big(x_j-x_i \big)},
\end{equation}
in which $\lb_i>0$ is the same quantity of Assumption \ref{ass.M_eps},   $k_i>0$ is a free control gain chosen   to satisfy   
\begin{equation}  \label{inq.ki_1} 
0< k_i \le  \dfrac{1}{\card([i]\setminus i)} 
\end{equation} 
and $h_i:\N\to \R_{\ge 0}$ is a time signal to be designed later on.

Notice that, as in \cite{nedic_constrained_2010}, the update laws  \eqref{s.xi} have the form of a projected (onto the interval $[\lb_i,\,\M_i]$) consensus-like protocol. Unlike \cite{nedic_constrained_2010}, however,  the matrix defining the estimates dynamics needs {\em not} be column or row-stochastic, and the coefficients $k_i$   are only constrained by \eqref{inq.ki_1} and, hence, they can be chosen in a completely decentralized way. Moreover, unlike all the aforementioned distributed optimization approaches, the restriction of the dynamics onto the \emph{consensus manifold}\footnote{That is, the set $\{x\in\R^N\st x_i=x_j,\ \forall i,j\in\cN\}$.} is not marginally stable. Rather, it is deliberately made unstable by the terms $\e^{h_i^t}$.

\subsection{Excitation Properties}

The signals $h_i$ will be chosen to guarantee one of the following \emph{excitation properties}.

\begin{definition}[Sufficiency of Excitation]\label{d.SE}
With $t_0\in\N$, the family $(h_i)_{i\in\cN}$, is said to be \emph{sufficiently exciting from $t_0$} if there exist $\und h(t_0)>0$ and $\Delta(t_0)\in\N_{\ge 1}$ such that, for each    $m\in\N_{\ge 1}$ satisfying
\begin{align}\label{e.SE.m}
 m &\le  \dfrac{1}{\und h(t_0)}\log\left( \dfrac{\uM}{\ulb} \right) 
\end{align}  
and each $i\in\cN$, there exists at least one     $s_i\in\{t_0+1+(m-1)\Delta(t_0),\,\dots,\,t_0+m\Delta(t_0)\}$ such that $h_i^{s_i}\ge \und h(t_0)$.
\end{definition}
In qualitative terms, given an initial time $t_0$, sufficiency of excitation  implies  that  the signals $h_i$ are positive  ``frequently enough" for a ``large enough" amount  of time succeeding $t_0$.
When $(h_i)_{i\in\cN}$ is sufficiently exciting from \emph{every} $t_0$, and independently on it, then we say that $(h_i)_{i\in\cN}$ enjoys the \emph{uniformity of excitation} property.
\begin{definition}[Uniformity of Excitation]\label{d.PE}
The family $(h_i)_{i\in\cN}$ is said to be \emph{uniformly exciting} if it is sufficiently exciting from every $t_0$,  with $\und h$ and $\Delta$ not dependent on $t_0$.  
\end{definition}
Uniformity of excitation can be seen as a ``uniform in $t_0$'' version of sufficiency of excitation and, in particular, it implies that all the signals $h_i$ take  positive values infinitely often. Defined in this way, both these properties are ``centralized'', in that they employ quantities common to all the agents. However, both can be easily obtained by means of decentralized design policies  in which the signals $h_i$ are chosen independently on each other. This is the case, for instance, when the signals $h_i$ are \emph{periodic} (with possibly different periods) and not identically zero, as formalized in the following lemma (proved in  \ref{apd.lemma_PE}).
\begin{lemma}\label{lem.PE}
Suppose that,  for each $i\in\cN$, $h_i$ is periodic and there exists $t\in\N$ for which $h_i^t >0$. Then, the family $(h_i)_{i\in\cN}$ is uniformly exciting.
\end{lemma} 

\begin{remark}
If $h_i^t=0$ for all $i\in\cN$ and $t\in\N$, each of the infinite points of the consensus manifold $\cM$ is an equilibrium for \eqref{s.xi}. Since $\uM\in\cM$, this implies that $\uM$ is a well-defined steady state for \eqref{s.xi}. However, in this case $\uM$ cannot be reached by any of the initial conditions in $\cM$, as they are indeed equilibria. This, in turn, is  related to the impossibility result \cite[Theorem 3.1.1]{Santoro2006} in the leader election problem in absence of unique identifiers, and is at the basis of the non-globality of the FloodMax and Max-Consensus algorithms (see Section \ref{sec.intro}).
In order to prevent the consensual states in $\cM$ to be equilibria, the signals $h_i^t$ must carry enough excitation, in the sense of Definitions \ref{d.SE} or \ref{d.PE}. As formally stated later on in Theorem \ref{thm.main}, indeed,   this permits to recover globality, although it ruins ``exactness'' of convergence of each estimate $x_i$ to $\uM$, being it a consensual state. In these terms, the signals $h_i$ play the same role of the \emph{dithering} signals in Extremum Seeking approaches \cite{Tan2006,Ariyur2003}.
\end{remark}

\section{Convergence Results}\label{sec:conv}
\subsection{Main result}

For ease of notation, we write the update laws \eqref{s.xi} in the compact form \eqref{s.x}.
The following theorem -- which is the main  result of the paper --  relates the excitation properties of the signals $h_i$  to  the asymptotic convergence  of the estimates $x_i$ produced by the update laws \eqref{s.xi} to $\uM$. In particular, it shows that sufficiency of excitation implies convergence (possibly exact) and uniformity of excitation implies uniform convergence, but ruins exactness.
  Further remarks and insights on the results given in the theorem follow  thereafter in Section~\ref{sec.remarks}.
\begin{theorem}\label{thm.main}
	Under Assumptions \ref{ass.connected} and \ref{ass.M_eps}, consider the update laws \eqref{s.xi}, in which $k_i$ satisfies \eqref{inq.ki_1}. Suppose that, for a given $t_0\in\N$, the family $(h_i)_{i\in\cN}$ is sufficiently exciting from $t_0$ in the sense of Definition \ref{d.SE}. Then, the   following claims  hold:
\begin{enumerate}
	\item \label{thm.main.item1} There exists $t\sr=t\sr(t_0)$  such that every solution $x$ to~\eqref{s.x} starting at $t_0$  satisfies
	\begin{equation*}
	\begin{array}{lclcl}
		x_i^t &\ge& \uM, && \forall t\ge t\sr(t_0),\ \forall i \in\cN\setminus I\sr\\
		x_i^t &=& \uM, && \forall t\ge t\sr(t_0),\ \forall i\in I\sr,
	\end{array}
	\end{equation*}
	with $I\sr$ given by \eqref{d.Isr}.

	\item \label{thm.main.item2} For each $\epsilon>0$, there exists $\delta(\epsilon)>0$ such that, if
	\begin{equation}\label{in.ls_hi}
	\limsup_{t\to\infty} h_i^t \le \delta(\epsilon),\quad \forall i\in\cN ,
	\end{equation}
	then each solution $x$ starting at $t_0$ satisfies   
	\begin{equation}\label{e.lim_xi_Ai}
	\lim_{t\to\infty}|x_i^t-\uM| \le \epsilon ,\quad \forall i\in\cN.
	\end{equation} 
	In particular, the set 
	\begin{equation*}
	\cA_\epsilon:= \prod_{i\in\cN} \big[\uM,\,\min\{\uM+\epsilon,\,\M_i\}\big]
	\end{equation*}
	 is globally attractive for \eqref{s.x}.  
	
	\item   \label{thm.main.item3}  If  the family $(h_i)_{i\in\cN}$ is uniformly exciting in the sense of Definition \ref{d.PE}, then $\cA_\epsilon$  is globally  uniformly   attractive. 
	
	\item If all the signals $h_i$ are non-zero and periodic (with possibly different periods), then there exists a compact set  $\cA_\epsilon^u\subset\cA_\epsilon$ which is globally  uniformly   attractive and stable, hence, globally uniformly asymptotically stable.
	\item \label{thm.main.item4} If   
	\begin{equation*}
	\lim_{t\to\infty} h_i^t =0,\quad \forall i\in\cN 
	\end{equation*}
	then, the set $\cA$, given by \eqref{d.A}, is globally attractive for \eqref{s.x}, i.e. 
	 \begin{equation*}
	\lim_{t\to\infty} x_i^t=\uM ,\quad\forall i\in\cN.
	\end{equation*}
\end{enumerate} 
\end{theorem}

For the reader's convenience, the proof of Theorem~\ref{thm.main} is postponed to Section \ref{sec.proof}.

\subsection{Remarks on the Result}\label{sec.remarks}

Claim 1 of Theorem \ref{thm.main} states that, if the family $(h_i)_{i\in\cN}$ is sufficiently exciting, then, in a finite time $t\sr$  the estimates $x_i$ of the agents $i\in I\sr$  satisfying $\M_i=\uM$  reach the target value $\uM$, while all the other estimates $x_i$ of the remaining agents $i\in\cN\setminus I\sr$ become larger than $\uM$. The time $t\sr$ is, however, a centralized quantity which depends on the excitation properties of all the signals $h_i$.

Claim 2  characterizes the asymptotic behavior of the remaining agents, by stating that the update laws \eqref{s.xi} are able to drive the estimates $x_i$  arbitrarily close to $\uM$, provided that  the amplitude of the signals $h_i^t$ is eventually reduced accordingly. 
As the approximation $\cA_\epsilon$ can be made arbitrarily tight, by acting on the asymptotic bounds of $h_i$ accordingly, it turns out that this is a  \emph{global practical attractiveness} result of the target set $\cA$ (defined in \eqref{d.A}). More precisely, let $\Gamma$ be the set of all the families   $\gamma:=(h_i)_{i\in\cN}$ of functions $h_i:\N\to\R_{\ge 0}$, and consider a  family of systems of the form \eqref{pre.s.xa}, with $x^t\in\R^N$ and $f_\gamma:=(f_{\gamma}^i)_{i\in\cN}$ satisfying
\begin{equation}\label{d.fgamma}
f_\gamma^i(t,x) := \proj{[\lb_i,\, \M_i]}{\e^{h_i^t} x_i + k_i \sum_{j\in[i]}\big(x_j-x_i \big)} .
\end{equation} 
Then, the second claim of the theorem can be restated as follows.

\begin{corollary}\label{cor.p}
Under the assumptions of Theorem \ref{thm.main}, the set $\cA$ is globally  practically attractive for the family \eqref{d.fgamma}.
\end{corollary}

Claim 3 of the theorem further strengthen Corollary~\ref{cor.p} to a \emph{uniform} global practical asymptotic stability property of $\cA$ in presence of uniformity of excitation. Moreover, in the relevant case in which the signals $h_i$ are periodic, Claim~4 guarantees the existence of a compact set included in   $\cA_\epsilon$ which is globally uniformly asymptotically stable.

Finally, Claim 5 states that, if all the signals $h_i^t$ converge to zero, then a \emph{global attractiveness} result of the target  set $\cA$ holds (i.e. $x_i^t \to \uM$ for all $i\in\cN$). However,  we observe that, if $h_i^t\to 0$ for some $i\in\cN$,  then the family $(h_i)_{i\in\cN}$ fails to be uniformly exciting, and thus the convergence  of the estimates $x_i$ to $\uM$ is \emph{not} in general uniform in the initial time $t_0$. This underlines an important difference between sufficiency and uniformity of excitation: sufficiency of excitation allows exact convergence, but prevents uniformity in the initial time.  Uniformity of excitation, instead, guarantees uniform convergence and  stability but frustrates exact convergence, guaranteeing only a weaker practical result. This, in turn, reveals a somehow   necessary compromise between complexity, uniformity and convergence.

\subsection{On the Design of the Signals $h_i$}
The signals $h_i$ are the only degrees of freedom  left to be chosen in the update laws \eqref{s.xi}. In this respect, Theorem \ref{thm.main} links their amplitude and excitation properties   to the corresponding asymptotic behavior of the estimates $x_i$, thus providing guidelines for their design. Based on the claims of Theorem \ref{thm.main}, in this section we discuss some possible designs guaranteeing sufficiency or uniformity of excitation.

\subsubsection{Sufficiently Exciting Designs}
Sufficiency of excitation of the family  $(h_i)_{i\in\cN}$ is guaranteed if each $h_i$ takes ``enough'' positive values. According to Definition \ref{d.SE}, and in particular to \eqref{e.SE.m}, how much is ``enough'' depends on centralized quantities. In turn, a design of the signals $h_i$ based on the knowledge of $t_0$ and of the quantities appearing in \eqref{e.SE.m} is undesirable as inevitably centralized  and not robust. A simple  decentralized way to design a sufficiently exciting family $(h_i)_{i\in\cN}$ amounts to choose bounded signals $h_i$ satisfying
\begin{equation}\label{e.sum_hi}
\sum_{t\in\N} h_i^t = \infty ,\qquad\forall i\in\cN.
\end{equation}
This, for instance, can be achieved by simply letting $h_i^t = a_i/(1+t)$ for some arbitrary $a_i>0$. 
\begin{lemma}\label{lem.SE}
Suppose that, for each $i\in\cN$, the signal $h_i$ is bounded and satisfies \eqref{e.sum_hi}. Then, the family $(h_i)_{i\in\cN}$ is sufficiently exciting in the sense of Definition \ref{d.SE}.
\end{lemma}
The proof of Lemma~\ref{lem.SE} follows directly from~\eqref{e.sum_hi}, hence it is omitted.

\smallskip

In view of Claim 5 of Theorem \ref{thm.main}, exact convergence of the estimates $x_i$ to $\uM$ is obtained if $\lim_{t\to\infty} h_i^t=0$ for all $i\in\cN$. Moreover, convergence of $h_i$ to zero is implied by (although not  equivalent to)  the following property
\begin{equation}\label{e.sum_hi2}
\sum_{t\in\N} \big(h_i^t\big)^2 < \infty .
\end{equation}
It is interesting to notice that Properties \eqref{e.sum_hi}-\eqref{e.sum_hi2} are  standard assumptions asked to the \emph{stepsize} in classical \emph{stochastic approximation algorithms} \cite{Robbins1951,Kushner1997}, as well as in modern distributed optimization algorithms using vanishing step sizes    \cite{NotarstefanoTutorial,nedic_constrained_2010,Simonetto2016}. In the context of this paper, these two conditions are simply sufficient conditions for sufficiency of excitation, which can be easily satisfied by decentralized designs of the signals $h_i$.

\subsubsection{Uniformly Exciting Designs}\label{sec.design_hi}
In view of Lemma \ref{d.PE}, if every signal $h_i$ is periodic, then $(h_i)_{i\in\cN}$ is uniformly exciting. While periodicity is not necessary for uniformity of excitation,   it certainly is  a relevant design choice due its simplicity and effectiveness.  Possible decentralized  design choices for periodic signals $h_i$ leading to a uniformly exciting family $(h_i)_{i\in\cN}$ are listed below, where the quantities $A_i,T_i,\rho_i>0$  are arbitrary. From the theoretical viewpoint, all the following options are equally fine. Depending on the application domain, however, some choices may be more convenient than others.
\begin{enumerate}
	\item \emph{Constant signals}:  is the simplest design choice and consists in choosing $h_i^t = A_i$ for all $i\in\cN$.
	\item \emph{Rectified sinusoids}: different versions can be defined, for instance $h_i^t = A_i |\sin(\pi t/T_i)|$ and $h_i=A_i\max\{0, \sin(2\pi t/T_i) \}$	both have period $T_i$.
	\item \emph{Square waves:} with $\rho_i\in(0,1]$ playing the role of a duty cycle, square waves have the form
	\begin{equation}\label{d.square_f}
	h_i^t = A_i {\rm step}\left({{\rm mod}(t,T_i)- (1-\rho_i)(T_i)}\right)
	\end{equation} 
	in which ${\rm mod}(s) := s-\max\{n\in\N \st n (T_i+1)\le s\}$, and ${\rm step}(\cdot)$ denotes the \emph{step function} satisfying ${\rm step}(s)=0$ for $s<0$ and ${\rm step}(s)=1$ for $s\ge 0$. The signal \eqref{d.square_f} has period $T_i$ and   $h_i^t=A_i$ holds for $\rho_iT_i$ seconds each period.
	\end{enumerate} 

\section{Numerical Simulations}\label{sec:simul}

\begin{figure}[h]
 	\vspace*{-.2cm}
	\centering
	\includegraphics[width=\linewidth,trim=5em 1em 5em 0em,clip]{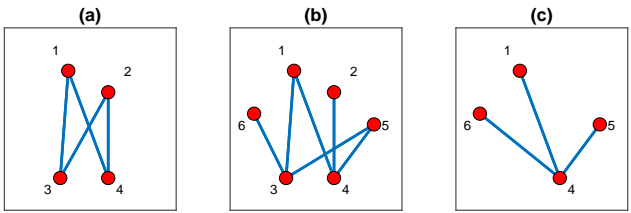}
	\vspace*{-.7cm} 
	\caption{Communication structure of Simulation 1:  \textbf{(a)} $[1]= \{1,3,4\}$, $[2]=\{2,3,4\}$, $[3]=\{1,2,3\}$, $[4]=\{1,2,4\}$; \textbf{(b)}  $[1]= \{1,3,4\}$, $[2]=\{2,4\}$, $[3]=\{1,3,5,6\}$, $[4]=\{1,2,4,5\}$, $[5]=\{3,4,5\}$ and $[6]=\{3,6\}$; \textbf{(c)}   $[1]=\{1,4\}$, $[4]=\{1,4,5,6\}$, $[5]=\{4,5\}$ and $[6]=\{4,6\}$.\vspace{-.3cm}}
	\label{Fig.ex1.top}
\end{figure}

\begin{figure*}
	\includegraphics[width=\linewidth,trim=4em 2em 4em 2em,clip]{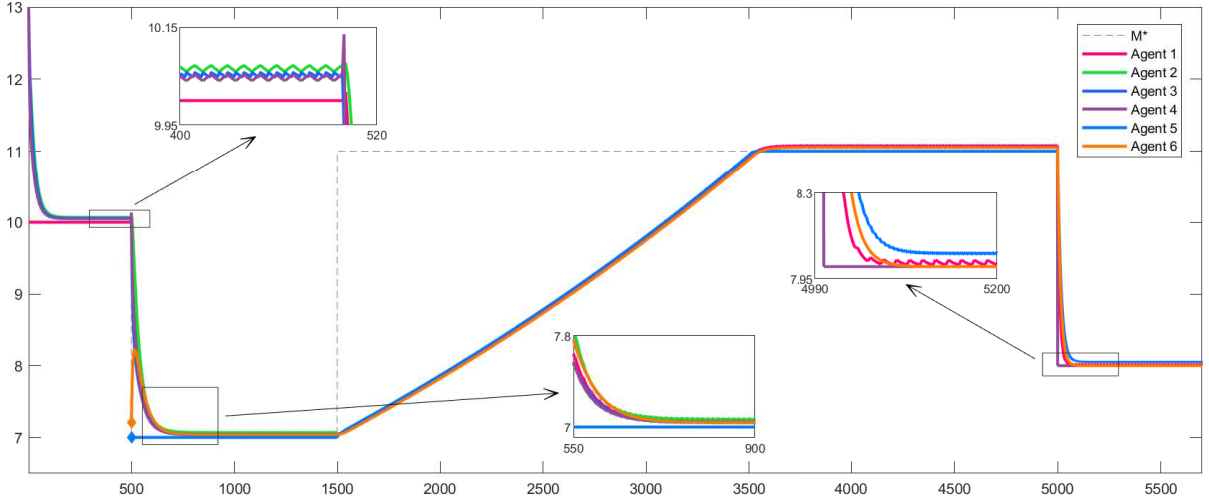}
	\vspace*{-.7cm}
	\caption{Evolution of the estimates $x_i$ in Scenario 1. The trajectory of the optimal value $\uM$ is shown in dashed gray line. Colored lines depict instead the trajectory of the estimates $x_i$, $i=1,\dots,6$. In abscissa: iteration variable $t$.\vspace{-.25cm}}
	\label{Fig.ex1.sim}
\end{figure*}

In this section, we present two illustrative numerical simulation scenarios. In Scenario~1, a network with a time-changing topology (see Figure~\ref{Fig.ex1.top}) is considered while in  Scenario~2, for a fixed network topology, the use of different signals $h_i$ is evaluated.

\subsection{Scenario 1: Uniform Convergence}

The first simulation, shown in Figure \ref{Fig.ex1.sim}, is obtained as follows.
The simulation starts with a network of $4$ agents (Agents $1$, $2$, $3$, and $4$), provided with a communication structure shown in Figure \ref{Fig.ex1.top}-(a)  and with   numbers $(\M_1,\,\M_2,\,\M_3,\,\M_4)=(10,\,12,\,13,\,13)$, implying $\uM=\M_1=10$.
	The update laws \eqref{s.xi} are implemented with $\lb_i=1/2$ for all $i\in\{1,\dots,4\}$, with $(k_1,\,k_2,\,k_3,\,k_4)=(0.1,\, 0.08,\,0.05,\,0.09)$, and with the signals $h_i$   chosen as the square waves discussed in Section \ref{sec.design_hi} with parameters $(T_1,A_1,\rho_1)=(15,10^{-3},0.2)$, $(T_2,A_2,\rho_2)=(10,5\cdot10^{-4},0.5)$, $(T_3,A_3,\rho_3)=(5, 10^{-3},0.3)$, $(T_4,A_4,\rho_4)=(10,5\cdot10^{-4},0.5)$.
	
At time $t=500$, two new agents (Agents $5$ and~$6$) are added to the network, and the communication structure is changed to the one shown in Figure~{\ref{Fig.ex1.top}-(b)}. The new agents have numbers $(\M_5,\M_6)=(7,11)$, lower bounds $\lb_5=\lb_6=1/2$, coefficients $(k_5,k_6)=(0.07,0.1)$, and signals $h_i$ given by the square waves presented in Section \ref{sec.design_hi} with $(T_5,A_5,\rho_5) = (5,10^{-3},0.4)$ and  $(T_6,A_6,\rho_6) = (7,25\cdot10^{-4},0.1)$. Furthermore, the  numbers of agents $1$ and~$3$ are changed to $(\M_1,\M_3)=(11,13)$. The new optimum is thus $\uM=\M_5=7$.
	
At time $t=1500$, Agents $2$ and $3$ leave the network, and the communication structure is changed to that depicted in Figure \ref{Fig.ex1.top}-(c). Moreover, the numbers of the agents are changed to $(\M_1,\M_4,\M_5,\M_6)=(12,16,11,16)$, leading to $\uM=\M_5=11$.
Finally, at time $t=5000$,  the number  of Agent  $4$ is changed to $\M_4=8$, so as $\uM=\M_4=8$.

As Figure \ref{Fig.ex1.sim} shows,  convergence to the (time-varying) optimum $\uM$ is approximate, and the trajectories of the agents   show residual oscillations. Figure~\ref{Fig.ex1.sim} also underlines that   convergence to $\uM$ ``from below'' (i.e. when the initial values of the agent are smaller than $\uM$) is slower than convergence ``from above'' (i.e. when the initial values of the agent are larger than $\uM$). As shown in the analysis of Section \ref{sec.proof}, this is due to the fact that (i)  the convergence rate ``from below'', proved in Section~\ref{sec.proof.1}, is determined by the   values of the signals $h_i^t$, while (ii) the convergence rate ``from above'', proved in Sections \ref{sec.proof.2}-\ref{sec.proof.3}, is determined by the   values of the coefficients $k_i$. 
\begin{figure}[h]
	\vspace*{-.25cm}
	\centering
	\includegraphics[width=\linewidth,trim=4em 0em 3em 0em,clip]{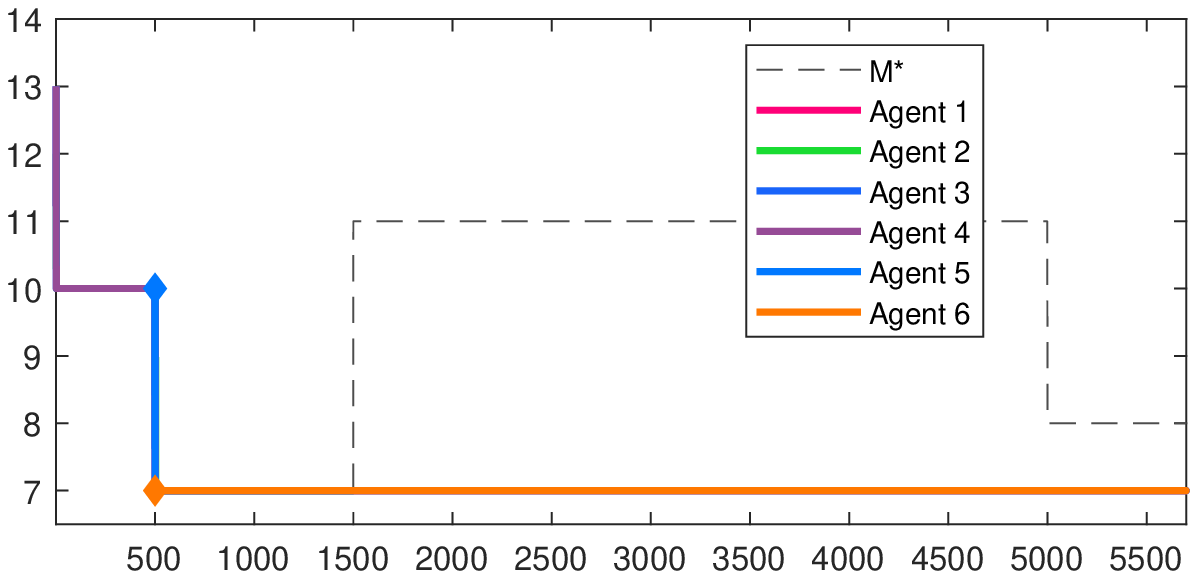}
	\vspace*{-.7cm}
	\caption{Evolution of the Max-Consensus estimates (update law~\eqref{s.ex.maxconsensus})   in the   setting of Scenario 1 (cf. Figure~\ref{Fig.ex1.sim}). In abscissa: iteration variable $t$.\vspace{-.25cm}}
	\label{Fig.maxc1}
\end{figure}

For the sake of comparison, Figure~\ref{Fig.maxc1} shows a simulation in which the Max-Consensus~\eqref{s.ex.maxconsensus} is employed in the same setting. As shown in Figure~\ref{Fig.maxc1},  although showing a faster convergence for the first two changes of $\uM$, the Max-Consensus fails in tracking the other changes. As illustrated in Section~\ref{sec.intro}, this is due to the fact that it is not globally attractive.

\subsection{Scenario 2: Non-Uniform Convergence}
\begin{figure*}
	\includegraphics[width=\linewidth,trim=4em .5em 4em 2em,clip]{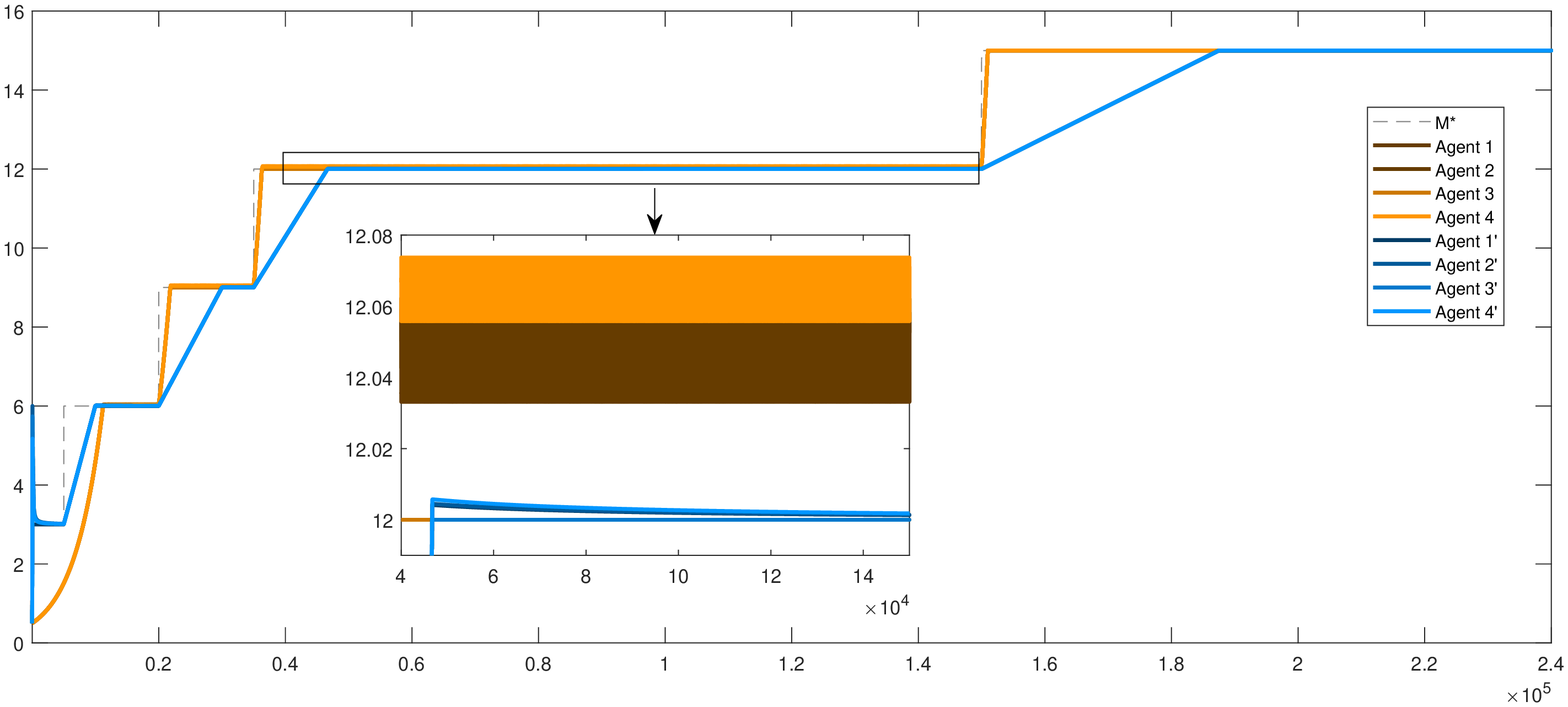}
	\vspace*{-.7cm}
	\caption{Evolution of the estimates $x_i$ in Scenario 2. The trajectory of the optimal value $\uM$ is shown in dashed gray line. Dark to light orange lines depict   the trajectory of the estimates $x_i$, $i=1,\dots,4$ of the first network. Dark to light blue lines depicts   the trajectory of the estimates $x_i$, $i=1',\dots,4'$ of the second network. In abscissa: iteration variable $t$.
		\vspace{-.45cm}}
	\label{Fig.ex2.sim}
\end{figure*}

\begin{figure}[h]
	\vspace*{-.1cm}
	\centering
	\includegraphics[width=\linewidth,trim=4em 0em 3em 0em,clip]{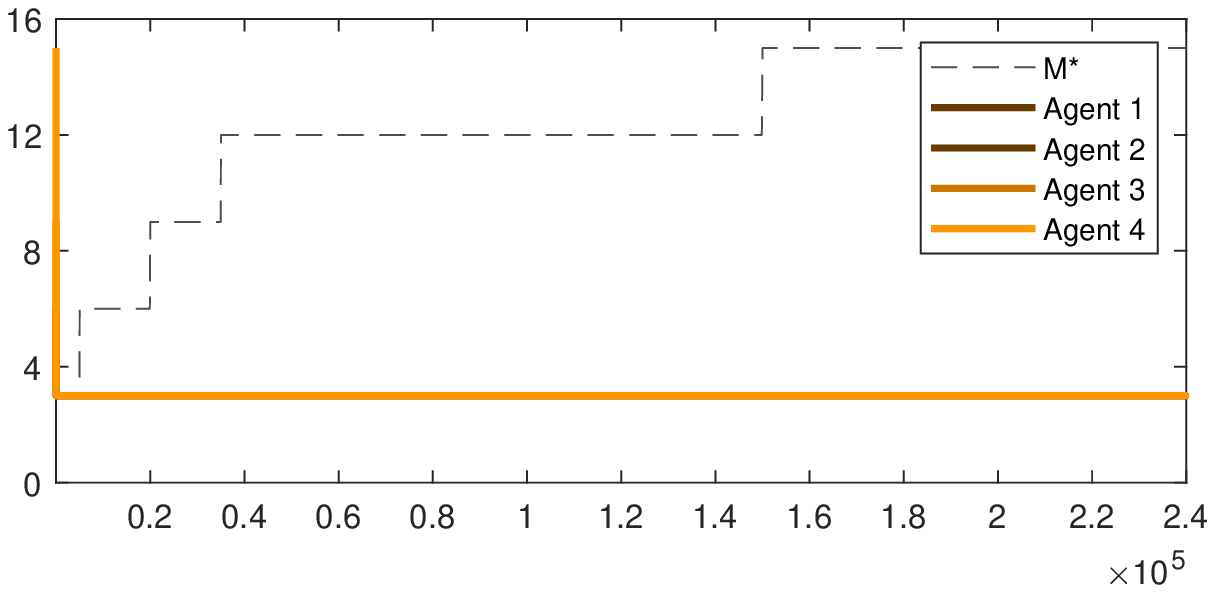}
	\vspace*{-.7cm}
	\caption{Evolution of the Max-Consensus estimates (update law~\eqref{s.ex.maxconsensus})   in the   setting of Scenario 2 (cf. Figure~\ref{Fig.ex2.sim}). In abscissa: iteration variable $t$.\vspace{-.5cm}}
	\label{Fig.maxc2}
\end{figure}

In the second scenario, we compare two simple networks having the same data and communication structures, but different signals $h_i$. The first network, $\cN$, includes Agents $1$, $2$, $3$ and $4$, and it is given the communication structure  depicted in Figure \ref{Fig.ex1.top}-(a). Initially, the agents are given numbers $(\M_1,\M_2,\M_3,\M_4)=(3,6,9,15)$, so as $\uM=\M_1=3$. At time $t=500$, $\M_1$ is changed to $15$, so as $\uM=\M_2=6$. At time $t=20000$, $\M_2$ is changed to $15$, so as $\uM=\M_3=9$. At time $t=35000$, $\M_3$ is changed to $12$, so as $\uM=M_3=12$. Finally, at time $t=150000$, $\M_3$ is changed to $15$, so as $\uM=\M_1=\M_2=\M_3=\M_4=15$. The update laws are implemented with $(k_1,k_2,k_3,k_4) = (0.1,0.08,0.05,0.09)$,  $\lb_1=\lb_2=\lb_3=\lb_4=1/2$, and with a family $(h_i)_{i\in\cN_1}$ of uniformly exciting signals defined as square waves with parameters $(T_1,A_1,\rho_1)=(15,10^{-3},0.2)$, $(T_2,A_2,\rho_2)=(10,5\cdot10^{-4},0.5)$, $(T_3,A_3,\rho_3)=(5, 10^{-3},0.3)$, $(T_4,A_4,\rho_4)=(10,5\cdot10^{-4},0.5)$.

The second network, $\cN'$, includes Agents $1'$, $2'$, $3'$ and $4'$ and has the same communication structure and data of~$\cN$. The update laws have the same parameters $k_{i'}=k_i$ and $\lb_{i'}=\lb_i$, $i\in\cN$, except for the family   $(h_{i'})_{i'\in\cN'}$ which is given by $h_{i'}^t =  (1+t)\inv$ for all $i'\in\cN'$. 
The signals $h_{i'}$ satisfy \eqref{e.sum_hi}-\eqref{e.sum_hi2} and, thus, $(h_{i'})_{i'\in\cN'}$ is sufficiently exciting. However, it fails to be uniformly exciting. The simulation  shown in Figure \ref{Fig.ex2.sim}  compares the time behavior of the update laws $x_i$, $i\in\cN$ and $x_{i'}$, $i'\in\cN'$. As shown in the figure, each ``step'' of $\uM$ is followed by the estimates $x_i$ with the same convergence rate. On the contrary, ${\uM}'=\uM$ is followed by the estimates $x_{i'}$ with a convergence rate which degrades in time. This is due to the fact that the family $(h_i)_{i\in\cN}$ is uniformly exciting, while the family $(h_{i'})_{i'\in\cN'}$ is only sufficiently exciting. Thus, uniformity of convergence   is not guaranteed for the estimates $x_{i'}$. Nevertheless, the zoomed part of the plot clearly shows that the estimates $x_{i'}$ reach ${\uM}'$ with higher precision (by Claim 5 of Theorem~\ref{thm.main},   indeed, since $h_{i'}^t\to 0$ the convergence of the estimates $x_{i'}$ is asymptotic if ${\uM}'$ remains constant), whereas the estimates $x_i$ exhibit a non-zero residual error. 
The above simulations underline the necessary compromise, already mentioned in different parts of the paper, and formally characterized by Claims 3  and 5 of Theorem \ref{thm.main}, between \emph{exact convergence} and \emph{uniformity in time}, which characterizes the proposed methodology.

Finally, Figure~\ref{Fig.maxc2} shows a simulation of the Max-Consensus \eqref{s.ex.maxconsensus} in the same   setting (cf. Figure~\ref{Fig.ex2.sim}). Again, the Max-Consensus fails in tracking the time-varying $\uM$. To see why this is the case, consider for instance the change of value of $\M_1$    at $t=500$. This determines an increment of $\uM$, bringing the  Max-Consensus algorithm   in a situation in which \eqref{s.ex.init2} holds at $t_0=500$. Hence, as explained in Section~\ref{sec.intro}, $x^{500}$ falls outside the domain of attraction of the new $\uM$, and thus convergence fails. 

\section{Concluding Remarks}\label{sec:concl}

As detailed in the proof of the main result (Section~\ref{sec.proof}) and shown in the numerical simulations, the proposed solution is characterized by a necessary compromise between convergence rate and asymptotic error, as both are determined in the worst case by the signals $h_i$. In particular, if $(h_i)_{i\in\cN}$ is uniformly exciting, uniform convergence is guaranteed, but the estimates will have a non-zero steady-state error. We stress that this residual error can be reduced arbitrarily by reducing the maximum value of the signals $h_i$ accordingly. But we also remark that, in general, this results in a reduction of the convergence rate. Larger values of the signals $h_i$ are associated instead with faster convergence but lead to larger steady-state errors. Moreover,
in the limit case in which $h_i^t\to 0$ for all $i\in\cN$, asymptotic convergence is obtained whenever $(h_i)_{i\in\cN}$ is sufficiently exciting. The convergence rate, however, is superlinear and not lower-bounded, and thus uniformity is lost. 

Clearly, ``smart'' choices of the signals $h_i$ are possible adapting their value at run time to increase them when fast convergence is needed and decrease them when, instead, we desire a low residual error. ``Adaptive'' design choices of this kind will be the subject of future research.

We prove all the proposed solution properties under the assumption that the communication structure and the parameters remain constant during the execution. 
Although uniform global asymptotic stability already guarantees a good behavior for ``slowly varying'' structures (also shown by the numerical simulations), additional work is needed to extend the analysis to handle time-varying networks with communication delays and noise. This extension, in turn, calls for a stochastic framework in which the aleatory nature of those phenomena is fully captured and is the subject of future research.

\section{Proof of Theorem \ref{thm.main}}  \label{sec.proof} 

\subsection{Proof of Claim 1}\label{sec.proof.1}
In this subsection we prove Claim 1.  In particular, we show that if the family $(h_i)_{i\in\cN}$ is sufficiently exciting from some $t_0\in\N$,  
then  there exists $t\sr=t\sr(t_0)>t_0$ such that, for each $i\in\cN$, $x_i^t\ge \uM$ holds for all $t\ge t \sr$ and,  for each $i\in I\sr$,    $x_i^t= \uM$ holds for all $t\ge t \sr$.

 Define the function $\und{i}:\R^N\to\cN$, $x\mapsto \und{i}(x):= \argmin_{i\in\cN} x_i$. 
Then, $x_j\ge x_{\und{i}(x)}$ holds for all $j\in\cN$. Moreover, $h_i^t\ge 0$ and   \eqref{inq.ki_1} imply $\e^{h_i^t}-\card([i]\setminus i)k_i \ge 0$ for all $i\in\cN$. 
Since $\projOp{[\lb_i,\M_i]}$ is increasing,  we have
\begin{equation}\label{pf.inq.xplus_1}
 \begin{aligned}
 x_i^{t+1} &= \proj{[\lb_i,\M_i]}{\left(\e^{h_i^t}-\card([i]\setminus i)k_i\right) x_i^t + k_i\sum_{j\in[i]\setminus i}  x_j^t }\\
 &\ge \projOp{[\lb_i,\M_i]}\bigg[\left(\e^{h_i^t}-\card([i]\setminus i) k_i\right) x_{\und i(x^t)}^t 
 \\&\hspace{4em}+ \card([i]\setminus i) k_i x_{\und i(x^t)}^t \bigg]\\
 &= \proj{[\lb_i,\M_i]}{\e^{h_i^t}x_{\und i(x^t)}^t } \\
 &= \max\left\{ \lb_i, \min\left\{ \e^{h_i^t}x_{\und i(x^t)}^t,\, \M_i \right\} \right\}\\
 &\ge   \min\left\{ \e^{h_i^t}x_{\und i(x^t)}^t,\, \M_i \right\} \ge \min\left\{ \e^{h_i^t}x_{\und i(x^t)}^t,\, \uM \right\} 
 \end{aligned}
\end{equation}
 for all $t\ge t_0$ and all $i\in\cN$.
 
First, notice that, if for some $\bar t\in\N$, $x_{\und i(x^{\bar t})}^{\bar t}\ge \uM$, then \eqref{pf.inq.xplus_1} implies  $x_{\und i(x^{\bar t+1})}^{\bar t+1}\ge \uM$, so that by induction it is possible to conclude that $x_i^t \ge \uM$ holds for all $t\ge\bar t$. Namely, the claim holds with $t\sr=\bar t$. It thus suffices to show that such  $\bar t$ exists. In doing so, we   proceed by contradiction. We first assume that 
\begin{equation}\label{pf.e.xontr}
x_{\und i(x^{t})}^{t}< \uM, \qquad \forall t\ge t_0.
\end{equation}
Then, we show  that, if the signals $h_i$ are sufficiently exciting from $t_0$ (in the sense of Definition \ref{d.SE}), then \eqref{pf.e.xontr}   leads to a contradiction, in this way proving the claim. 

Thus, assume that \eqref{pf.e.xontr} holds.  Then, since $h_i^t\ge 0$ for all $i\in\cN$, \eqref{pf.inq.xplus_1} yields 
\begin{equation}\label{pf.in.xplus_2}
x_i^{t+s}\ge \e^{h_i^t}x_{\und i(x^t)}^t,\qquad\forall t \ge t_0, \ s\ge 1.
\end{equation}
Suppose that the signals $h_i$ are sufficiently exciting from $t_0$, for some parameters $\und h(t_0)$ and $\Delta(t_0)$. Then, for each $i\in\cN$, there exists $s_i\in\{t_0+1,\,\dots,\,t_0+\Delta(t_0)\}$, such that $h_i^{s_i}\ge \und h(t_0)$. In view of \eqref{pf.in.xplus_2}, this yields
\begin{equation*}
x^{t_0+1+\Delta(t_0)}_i \ge \e^{\und h(t_0)}\, x_{\und i(x^{t_0+1})}^{t_0+1},\qquad\forall i\in\cN ,
\end{equation*}
and thus, in particular, 
\begin{equation*}
x^{t_0+1+\Delta(t_0)}_{\und i\big(x^{t_0+1+\Delta(t_0)}\big)} \ge \e^{\und h(t_0)}\, x_{\und i(x^{t_0+1})}^{t_0+1}  .
\end{equation*}
In the same way,   in view of   sufficiency of excitation of the signals $h_i$, for each $i\in\cN$, there exists $s_i\in\{t_0+1+\Delta(t_0),\,\dots,\,t_0+2\Delta(t_0)\}$, such that $h_i^{s_i}\ge \und h(t_0)$. Then, in view of \eqref{pf.in.xplus_2}, one has
\begin{align*}
x^{t_0+1+2\Delta(t_0)}_{\und i\big(x^{t_0+1+2\Delta(t_0)}\big)}  \ge \e^{\und h(t_0)}\, x^{t_0+1+\Delta(t_0)}_{\und i\big(x^{t_0+1+\Delta(t_0)}\big)} \ge  \e^{2\und h(t_0)}\, x_{\und i(x^{t_0+1})}^{t_0+1}.
\end{align*}
By repeating the same arguments,  it is thus possible to conclude that, for each $m\in\N$ satisfying \eqref{e.SE.m}, one has
\begin{equation}\label{pf.in.xplus_3}
x^{t_0+1+m\Delta(t_0)}_{\und i\big(x^{t_0+1+m\Delta(t_0)}\big)} \ge \e^{m\und h(t_0)}\, x_{\und i(x^{t_0+1})}^{t_0+1} \ge  \e^{m\und h(t_0)} \ulb ,
\end{equation} 
in which we used the fact that, by definition of $\projOp{[\ulb_i,\M_i]}$, $x_i^t\ge \lb_i\ge\ulb$ for all $i\in\cN$ and all $t\ge t_0+1$.
 Since the latter relation holds in particular for
 \begin{equation*}
 m\sr(t_0) = \dfrac{1}{\und h(t_0)}\log\left( \dfrac{\uM}{\ulb} \right) .
 \end{equation*}
 Then, with $\bar t:= t_0+1+m\sr(t_0)\Delta(t_0)$, from \eqref{pf.in.xplus_3} we obtain
 \begin{align*}
 x_i^{\bar t} \ge x^{\bar t}_{\und i(x^{\bar t})} \ge \e^{m\sr(t_0)\und h(t_0)} \ulb = \uM ,\qquad\forall i\in\cN
 \end{align*}
 which contradicts \eqref{pf.e.xontr} and, thus, proves that $x_i^t \ge \uM$ holds for all $i\in\cN$ and all $t\ge t\sr:=\bar t$.
 
 Finally, for all $i\in I\sr$, we have $x_i^t\in[\ulb,M_i]\le \uM$   for all $t\ge t_0+1$ and this, together with the bound   $x_i^t\ge \uM$ above, implies $x_i^t=\uM$  for all $i\in I\sr$ and $t\ge t\sr$.

\subsection{Proof of Claim 2}\label{sec.proof.2}

Since by Claim 1 each $x_i$ satisfies $x_i^t\ge \uM$ for all $t\ge t\sr$, then, in view of  Assumption \ref{ass.M_eps}, each $x_i$ also satisfies $x_i^t\ge \lb_i$ for all $t\ge t\sr$. This, in turn, allows us to write  
\begin{equation*}
x_i^{t+1} = \min\left\{ \M_i,\ \e^{h_i^t} x_i^t + k_i \sum_{j\in[i]} \big(x_j^t-x_i^t\big) \right\}
\end{equation*}
for all $i\in\cN$ and all $t\ge t\sr$, which  implies both
\begin{equation}\label{pf.e.xi_le_Mi}
x_i^{t} \le \M_i  
\end{equation}
and
\begin{equation}\label{pf.s.xi_le}
x_i^{t+1} \le   \e^{h_i^t} x_i^t + k_i \sum_{j\in[i]} \big(x_j^t-x_i^t\big)
\end{equation}
for all $i\in\cN$ and all $t\ge t\sr$.
From \eqref{pf.e.xi_le_Mi} we also obtain
\begin{equation}\label{pf.ineq.limsup_1}
\limsup_{t\to\infty} |x_i^t| \le \M_i <\infty,\qquad \forall i\in\cN.
\end{equation}
In the following we rely on the forthcoming lemma, whose proof is postponed to \ref{apd.Lemma_limsup}.
\begin{lemma}\label{lem.limsup}
	With $n\in\N$, let $x,\,y:\N\to \R^n$. Suppose that $y$ is  {bounded} and that, for some $t_0\in\N$ and some $\lambda:\N\to\Rplus$ fulfilling $\lambda^t \le  \nu \in [0,1)$ for all $t\ge t_0$, $x$ and $y$ satisfy	
	\begin{equation}\label{e.lem.xy}
		x^{t+1} \le \lambda^t x^t + y^t  
	\end{equation}
	  for all $t\ge t_0$. Then
	\begin{equation}\label{pf.s.ls_xi_le}
		\limsup_{t\to\infty}|x^t|\le  \dfrac{1}{1-\limsup_{t\to\infty}\lambda^t}\limsup_{t\to\infty}|y^t|.
	\end{equation} 
\end{lemma}

With $I\sr$ defined in \eqref{d.Isr}, let $n\sr$ be the least integer such that $[I\sr]^{n\sr}=\cN$ (which exists finite in view of Assumption~\ref{ass.connected}). The case in which $n\sr=0$ (i.e. $I\sr=\cN$) directly follows from Claim 1. Hence, we consider $n\sr>0$. 

Assume that, for some $m\in\{0,\dots, n\sr-1 \}$, there exist  $\alpha_m\in[0,1)$  and  $\beta_m>0$ such that\footnote{Here we let $[I\sr]^{-1}:=\emptyset$.}
	\begin{equation}\label{pf.in.induct0}
	\begin{aligned}
	&\max_{i\in   [I\sr]^{m}_{m-1}}\limsup_{t\to\infty}|x_i^t|   \le   \alpha_m \max_{j\in [I\sr]^{m+1}_{m}} \limsup_{t\to\infty}|x_j^t|   + \beta_m \uM .
	\end{aligned}
	\end{equation} 
We will now prove that, if this is the case, then a similar property  holds also for $m+1$. 

First  notice that, for each $i\in [I\sr]^{m+1}_{m}$, every $j\in[i]$ belongs to exactly one among the sets   $[I\sr]^{m+2}_{m+1}$,   $[I\sr]^{m+1}_{m}$, and $[I\sr]^{m}_{m-1}$. Hence, in view of \eqref{pf.s.xi_le}, we can write
\begin{equation}\label{pf.in.xi_1}
\begin{aligned}
x_i^{t+1} & \le\big(\e^{h_i^t} - k_i \card([i]\setminus i)\big)x_i^t + k_i\sum_{j\in [i]\cap [I\sr]^{m} } x_j^t  \\&\qquad  + k_i\sum_{j\in ([i]\setminus i)\cap  [I\sr]^{m+1}_{m}} x_j^t  + k_i\sum_{j\in [i]\cap [I\sr]^{m+2}_{m+1}} x_j^t
\end{aligned}
\end{equation}
for all $i\in[I\sr]^{m+1}_{m}$ and all $t\ge t\sr$, in which we used the fact that $[i]\cap   [I\sr]^{m}_{m-1}  =  [i]\cap   [I\sr]^{m}$, for all $i\in[I\sr]^{m+1}_{m}$. If \eqref{inq.ki_1} holds, then $1+k_i \card([i]\setminus i)>1$. With $\nu_1>0$ sufficiently small so that $ \log(1+k_i \card([i]\setminus i))-2\nu_1> 0$, let 
\begin{equation*}
\bar h_{i,1} := \log(1+k_i \card([i]\setminus i)) -2\nu_1.
\end{equation*}
If $\limsup_{t\to\infty} h_i^t \le \bar h_{i,1}$ 
for all $i\in\cN$, then there exists $T\sr>t\sr$ such that
\begin{equation}\label{pf.in.barh}
h_i^t \le \bar h_{i,1}+\nu_1 = \log(1+k_i \card([i]\setminus i)) - \nu_1
\end{equation}
for all $t\ge T\sr$ and all $i\in\cN$.  Thus,  \eqref{inq.ki_1} and \eqref{pf.in.barh}  imply
\begin{equation*}
0\le \e^{h_i^t} - k_i  \card([i]\setminus i) \le \e^{\bar h_{i,1}+\nu_1}- k_i \card([i]\setminus i)  < 1,
\end{equation*}
for all $t\ge T\sr$ and all $i\in\cN$, 
so that \eqref{pf.ineq.limsup_1}, \eqref{pf.in.xi_1} and Lemma~\ref{lem.limsup} imply
\begin{equation}\label{pf.in.xi_2}
\begin{aligned}
\limsup_{t\to\infty} |x_i^{t}| & \le  \gamma_{i} \sum_{j\in [i]\cap [I\sr]^{m}} \limsup_{t\to\infty}|x_j^t| \\
&\qquad + \gamma_{i}\sum_{j\in ([i]\setminus i)\cap [I\sr]^{m+1}_{m}} \limsup_{t\to\infty}|x_j^t| \\&\qquad +\gamma_{i}\sum_{j\in [i]\cap [I\sr]^{m+2}_{m+1}} \limsup_{t\to\infty}|x_j^t|
\end{aligned}
\end{equation}
for all $i\in [I\sr]^{m+1}_m$, in which we let
\begin{equation}\label{pf.d.gammai}
\gamma_i :=  \dfrac{k_i}{1- \limsup_{t\to\infty}\,  \big(\e^{h_i^t} - k_i  \card([i]\setminus i)\big)}
\end{equation}
which exists 
finite in view of Lemma \ref{lem.limsup}. 
In view of \eqref{pf.in.induct0}, equation  \eqref{pf.in.xi_2} implies
\begin{equation}\label{pf.in.xi_3}
\begin{aligned}
\limsup_{t\to\infty} |x_i^{t}| & \le  
\big( c_{i,1} \alpha_m +c_{i,2}\big) \max_{j\in [I\sr]^{m+1}_{m}} \limsup_{t\to\infty}|x_j^t|\\ 
&\qquad+ c_{i,3} \max_{j\in [I\sr]^{m+2}_{m+1}} \limsup_{t\to\infty}|x_j^t|+ c_{i,1}    \beta_m \uM.
\end{aligned}
\end{equation}
for all $i\in [I\sr]^{m+1}_m$, in which we let for convenience
\begin{equation}\label{pf.d.ci}
\begin{aligned}
c_{i,1} &:= \gamma_i \card\left( [i] \cap [I\sr]^{m}\right)  \\
c_{i,2} &:= \gamma_i \card\left(([i]\setminus i)\cap [I\sr]^{m+1}_{m}\right)  \\
c_{i,3} &:= \gamma_i \card\left([i]\cap [I\sr]^{m+2}_{m+1}\right)  .
\end{aligned}
\end{equation}
With $\nu_2>0$ sufficiently small so that $k_i(1-\alpha_m)- \nu_2>0$ for all $i\in\cN$ (recall that $\alpha_m<1$ by assumption), define
\begin{equation*}
\bar h_i := \min\Big\{ \bar h_{i,1},\ \log\big(1+k_i(1-\alpha_m)- \nu_2\big)   \Big\}.
\end{equation*}
If
\begin{equation}\label{pf.in.ls_hi}
\limsup_{t\to\infty} h_i^t \le \bar h_i
\end{equation}
for all $i\in [I\sr]^{m+1}_m$, then, 
since $\card([i]\cap  [I\sr]^{m} )\ge 1$,   it holds that 
\begin{equation}\label{pf.in.hsr_0}
\begin{aligned}
1-\e^{\limsup_{t\to\infty} h_i^t} & \ge 1-\e^{\bar h_i} \ge  -k_i(1-\alpha_m)+\nu_2 \\
&\ge - k_i(1-\alpha_m)\card([i]\cap [I\sr]^{m}) + \nu_2  
\end{aligned}
\end{equation}
for all   $i\in [I\sr]^{m+1}_m$. 
Since for all $i\in [I\sr]^{m+1}_m$,
\begin{align*}
&\card\big(([i]\setminus i)\cap [I\sr]^{m+1}_{m}\big)\\
&= \card\left( [i]\setminus i\right) -\card\left( [i] \cap [I\sr]^{m}\right) - \card\left([i]\cap [I\sr]^{m+2}_{m+1}\right) \\
&\le  \card\left( [i]\setminus i\right) -\card\left( [i] \cap [I\sr]^{m}\right),
\end{align*}
then, we conclude that
\begin{equation}\label{pf.in.cis}
\begin{aligned}
c_{i,1}& \alpha_m + c_{i,2} \\
&\le \dfrac{ k_i(\alpha_m-1) \card\left( [i] \cap [I\sr]^{m}\right)+k_i \card\left( [i]\setminus i\right)}{1-\e^{\bar h_i}+ k_i \card([i]\setminus i) }\\
&\le \dfrac{(\alpha_m-1) k_i \card\left( [i] \cap  [I\sr]^{m}\right) +k_i \card([i]\setminus i)}{(\alpha_m-1) k_i \card\left( [i] \cap [I\sr]^{m}\right) +k_i \card([i]\setminus i)+\nu_2 }\\&<1.
\end{aligned}
\end{equation}
for all $i\in [I\sr]^{m+1}_m$.

Now, since \eqref{pf.in.xi_3} holds for each  $i\in [I\sr]^{m+1}_m$, it in particular holds for  $\bar i$ satisfying
\begin{equation}\label{pf.d.bari}
\bar i\in \argmax_{i\in [I\sr]^{m+1}_m} \limsup_{t\to\infty}|x_i^t|,
\end{equation} 
so that \eqref{pf.in.xi_3}  implies
\begin{equation*}
\begin{aligned}
\max_{i\in [I\sr]^{m+1}_{m}}& \limsup_{t\to\infty}|x_i^t|   \le  
 ( c_{\bar i,1} \alpha_m +c_{\bar i,2} ) \max_{i\in [I\sr]^{m+1}_{m}} \limsup_{t\to\infty}|x_i^t|\\ 
&\qquad\qquad+ c_{\bar i,3} \max_{j\in [I\sr]^{m+2}_{m+1}} \limsup_{t\to\infty}|x_j^t| + c_{\bar i,1}\beta_m \uM
\end{aligned}
\end{equation*}
which, in view of \eqref{pf.in.cis}, yields
\begin{equation}\label{pf.in.induct1}  
\begin{aligned}
\max_{i\in [I\sr]^{m+1}_{m}} \limsup_{t\to\infty}|x_i^t|   &\le  \alpha_{m+1} \max_{j\in [I\sr]^{m+2}_{m+1}} \limsup_{t\to\infty}|x_j^t| \\&\qquad +  \beta_{m+1} \uM 
\end{aligned}
\end{equation}
with
\begin{equation}\label{pf.d.a_b}
\begin{aligned}
\alpha_{m+1} &= \dfrac{c_{\bar i,3}}{1-\big( c_{\bar i,1} \alpha_m +c_{\bar i,2}\big)}, \\
\beta_{m+1} &=\dfrac{c_{\bar i,1}}{1-\big( c_{\bar i,1} \alpha_m +c_{\bar i,2}\big)} \beta_m .
\end{aligned}
\end{equation}
Furthermore,  since $\limsup_{t\to\infty} h_i^t\le \bar h_i$,   in view of \eqref{pf.in.hsr_0},   $\alpha_{m+1}$   satisfies 
\begin{align*}
\alpha_{m+1} &\le  \dfrac{k_{\bar i} \card \big( [{\bar i}]\cap [I\sr]^{m+2}_{m+1} \big) }{  k_{\bar i}\card([{\bar i}]\setminus {\bar i})-  k_{\bar i}  \card(([{\bar i}]\setminus {\bar i})\cap[I\sr]^{m+1})+\nu_2 }\\
&\le  \dfrac{k_{\bar i} \card \big( [{\bar i}]\cap [I\sr]^{m+2}_{m+1} \big) }{  k_{\bar i} \card \big( [{\bar i}]\cap [I\sr]^{m+2}_{m+1} \big)+\nu_2 } <1.
\end{align*} 
Therefore,  we claim that if \eqref{pf.in.induct0} holds for some $m\in\{0,\dots, n\sr-1\}$ with $\alpha_m<1$ and $\beta_m\ge 0$, then \eqref{pf.in.induct1} holds as well for $m+1$ with $\alpha_{m+1}<1$ and $\beta_{m+1}\ge 0$ given above.
Since  by Claim 1, Equation \eqref{pf.in.induct0} trivially holds for $m=0$ with $\beta_0=1$ and $\alpha_0=0$, then  we claim by induction that, if 
\begin{equation}\label{pf.d.barh}
\limsup_{t\to\infty} h_i^t \le \bar h:= \min_{i\in\cN} \bar h_i,\qquad \forall i\in\cN,
\end{equation}
then Equation  \eqref{pf.in.induct0} holds  for each $m\in\{0,\dots,n\sr\}$. 

Now, for $m=n\sr$, we have  $[I\sr]^{m+1}\setminus [I\sr]^m = \emptyset$, so that \eqref{pf.in.induct0} yields
\begin{equation*}
\limsup_{t\to\infty} x_i^t \le \beta_{n\sr}\uM , \qquad \forall i\in [I\sr]^{n\sr}_{n\sr-1}.
\end{equation*}
Thus, iterating \eqref{pf.in.induct0} backwards and using \eqref{pf.e.xi_le_Mi} yield 
\begin{equation}\label{ps.LS_xi}
\limsup_{t\to\infty}x_i^t \le \min\Big\{ \M_i,\ (1+\varepsilon_i)\uM  \Big\}
\end{equation}
in which
\begin{equation*}
\varepsilon_i  = 0,\qquad \forall i\in I\sr
\end{equation*}
 and
\begin{equation}\label{pf.d.vep_i} 
\varepsilon_i =   \sum_{\ell=0}^{n\sr-m} \left( \prod_{k=\ell+1}^{n\sr-m} \alpha_{n\sr-k} \right) \beta_{n\sr-\ell}
-1,  
\end{equation}
for all $i\in [I\sr]^{m}_{m-1} $ and all $m=1,\dots,n\sr$.  Moreover, \eqref{pf.d.vep_i} directly implies that the quantities $\varepsilon_i$ also satisfy  
\begin{equation}\label{pf.d.vep_i_2}
	\max_{i\in[I\sr]^m_{m-1}} \varepsilon_i = \alpha_m \left(1+\max_{i\in[I\sr]^{m+1}_{m}}\varepsilon_i\right) + \beta_m -1
\end{equation}
for all $m=1,\dots,n\sr$.
We now prove that $\varepsilon_i$ in \eqref{ps.LS_xi}-\eqref{pf.d.vep_i}  can be reduced arbitrarily by reducing $\limsup_{t\to\infty} h_i^t$   accordingly for each $i\in\cN$. For convenience, let
\begin{equation}\label{pf.d.upsilon}
\upsilon_i := \limsup_{t\to\infty} h_i^t \in [0,\bar h_i].
\end{equation}
Then, the quantities $\gamma_i$, defined in \eqref{pf.d.gammai}, satisfy 
\begin{equation*}
\gamma_i(\upsilon_i) = \dfrac{k_i}{1-\e^{\upsilon_i} + k_i\card([i]\setminus i)}.
\end{equation*}
Thus, $\gamma_i$ is continuous in $[0,\infty)$, and
\begin{equation*}
\lim_{\upsilon_i\to 0} \gamma_i(\upsilon_i) =  \dfrac{1}{\card([i]\setminus i)}.
\end{equation*}
In view of the definitions \eqref{pf.d.ci}, also the quantities $\alpha_m$ and $\beta_m$, as defined in \eqref{pf.d.a_b}, depend on $\upsilon_{\bar i}$ through $\gamma_{\bar i}$, in which $\bar i$ satisfies \eqref{pf.d.bari}. We now prove by induction that, by letting $\upsilon:=(\upsilon_1,\dots,\upsilon_N)$, the following holds
\begin{equation}\label{pf.e.lim_ab_1}
\lim_{\upsilon\to 0} \alpha_m(\upsilon)+\beta_m(\upsilon) = 1,\qquad\forall m=0,\dots,n\sr.
\end{equation}
First notice that \eqref{pf.e.lim_ab_1} trivially holds for $m=0$, as indeed $\alpha_m=0$ and $\beta_m=1$ despite the value of $\upsilon$. It thus suffices to show that if \eqref{pf.e.lim_ab_1} holds for a given $m\in\{0,\dots,n\sr-1\}$, then the same relation holds as well for $m+1$. For, assume that \eqref{pf.e.lim_ab_1} holds for a given $m\in\{0,\dots,n\sr-1\}$. Then, we can write $\lim_{\upsilon\to 0} \beta_m(\upsilon)=1-\lim_{\upsilon\to 0}\alpha_m(\upsilon)$. Thus, by letting for convenience $\rho_1:=\card([\bar i]\cap[I\sr]^m)$, $\rho_2:=\card(([\bar i]\setminus \bar i)\cap ([I\sr]^{m+1}_m))$, $\rho_3:=\card([\bar i]\cap([I\sr]^{m+2}_{m+1}))$, and noting that $\card([\bar i]\setminus \bar i)-\rho_2 = \rho_1+\rho_3$, we obtain
\begin{equation*}
\begin{aligned}
&\lim_{\upsilon\to 0} \alpha_{m+1}(\upsilon)+ \beta_{m+1}(\upsilon)  \\&\quad = \dfrac{\rho_3 + \left(1-\lim_{\upsilon\to 0}\alpha_m(\upsilon)\right) \rho_1}{\card([\bar i]\setminus \bar i) - \lim_{\upsilon\to 0}\alpha_m(\upsilon)\rho_1 - \rho_2}\\&\quad = \dfrac{\rho_3 + \left(1-\lim_{\upsilon\to 0}\alpha_m(\upsilon)\right) \rho_1}{\rho_3 + \left(1-\lim_{\upsilon\to 0}\alpha_m(\upsilon)\right) \rho_1}=1.
\end{aligned}
\end{equation*}
Thus, by induction, we claim \eqref{pf.e.lim_ab_1} for all $m\in\{0,\dots,n\sr\}$.

Since for every $i\in[I\sr]^{n\sr}_{n\sr-1}$, $c_{i,3} = 0$ (in fact $[I\sr]^{n\sr+1}_{n\sr}=\emptyset$), then $\alpha_{n\sr}=0$. Thus,
\begin{equation*}
\lim_{\upsilon\to 0}\beta_{n\sr}(\upsilon)=1.
\end{equation*}
In view of \eqref{pf.d.vep_i}, this implies
\begin{equation*}
 \lim_{\upsilon\to 0}\max_{i\in[I\sr]_{n\sr}^{n\sr-1}} \varepsilon_{i}(\upsilon) = 0.
\end{equation*}
In view of \eqref{pf.d.vep_i_2}, $\lim_{\upsilon\to 0}\max_{i\in[I\sr]_{m}^{m+1}}\varepsilon_i(\upsilon)=0$ implies
\begin{align*}
\lim_{\upsilon\to 0} \max_{i\in[I\sr]_{m-1}^{m}}\varepsilon_{i}(\upsilon) = \lim_{\upsilon\to 0} (\alpha_{m}(\upsilon)+\beta_m(\upsilon)) -1 = 0,
\end{align*}
so that, by induction, we conclude that
\begin{equation*}
\lim_{\upsilon\to 0} \max_{i\in[I\sr]_{m}^{m-1}}\varepsilon_i(\upsilon) = 0,\quad \forall m\in\{0,\dots,n\sr\},
\end{equation*}
i.e.
\begin{equation}\label{pf.eq.lim_vep}
\lim_{\upsilon\to 0}  \varepsilon_i(\upsilon) = 0,\quad \forall i\in\cN.
\end{equation}
The latter equation thus implies that, given any $\epsilon\ge0$, there exists $\delta'(\epsilon)\ge 0$ such that $|\upsilon|\le \delta'(\epsilon)$ implies $\uM \varepsilon_i\le \epsilon$ for all $i\in\cN$. Therefore, if
\begin{equation}\label{pf.inq.ls_hi}
\limsup_{t\to\infty}h_i^t \le \delta(\epsilon) := \min\left\{\bar h,\,\dfrac{\delta'(\epsilon)}{N}\right\},\qquad\forall i\in\cN 
\end{equation}
then $|\upsilon|\le \delta'(\epsilon)$, which implies $\uM\varepsilon_i\le \epsilon$. In turn, in view of \eqref{ps.LS_xi}, this implies 
\begin{equation}\label{ps.LS_xi2}
\limsup_{t\to\infty}x_i^t \le \min\Big\{ \M_i,\  \uM +\epsilon \Big\}.
\end{equation}
Claim 2 thus follows from \eqref{ps.LS_xi2} and by noticing that Claim~1 implies $\limsup_{t\to\infty} x_i\ge \uM$. 

\subsection{Proof of Claim 3}\label{sec.proof.3}
The third claim of the theorem, i.e., that uniformity of excitation (in the sense of Definition \ref{d.PE}) of $(h_i)_{i\in\cN}$ implies uniform attractiveness of $\cA_\epsilon:= \prod_{i\in\cN} \big[\uM,\,\min\{\uM+\epsilon,\,\M_i\}\big]$,  directly follows by the fact  that, if the family $(h_i)_{i\in\cN}$ is uniformly exciting,   then in the above analysis   $t\sr$ does not  depend on $t_0$ and, therefore,    the convergence \eqref{ps.LS_xi2} is uniform in the initial time.  

\subsection{Proof of Claim 4}
In this subsection we prove the fourth claim of the theorem.
With $(\tau_i)_{i\in\cN}\in\N^N$   arbitrary,  let $F_i\in\R^{\tau_i\x \tau_i}$ and $C_i\in\R^{1\x \tau_i}$ denote the matrices
\begin{align*}
F_i &:= \begin{bmatrix}
0_{(\tau_i-1)\x 1 }  & I_{(\tau_i-1)\x (\tau_i-1)}\\
1 & 0_{1\x(\tau_i-1)}
\end{bmatrix}, & C_i&:= \begin{bmatrix}
1 & 0_{1\x(\tau_i-1)}
\end{bmatrix}.
\end{align*}
Then, each $\tau_i$-periodic signal $h_i$   satisfies
\begin{equation}\label{pf.s.xxi}
\begin{aligned}
\xi^{t+1}_i &=  F_i\xi^t_i, & 
h^t_i &= C_i \xi^t_i
\end{aligned}
\end{equation}
for a suitable initial condition $\xi^{t_0}_i\in\R^{\tau_i}$. Moreover, if all the signals $h_i$ are non-zero, then, by Lemma \ref{lem.PE}, $(h_i)_{i\in\cN}$ is uniformly exciting in the sense of Definition \ref{d.PE} for some    $\und h>0$. For a fixed $\epsilon>0$, let $\delta(\epsilon)$ be defined as above in~\eqref{pf.inq.ls_hi}, and let
\begin{align*}
\Xi_i :=\Big\{ \xi_i\in\R^{\tau_i} \st\, & \forall j\in\{1,\dots,\tau_i\},\, \xi_{i,j}\in[0,\delta(\epsilon)],\text{ and }\\
& \exists j\in\{1,\dots,  \tau_i\},\, \xi_{i,j}\ge \und h\Big\},
\end{align*}
 where  $\xi_{i,j}$ denotes the $j$-th component of $\xi_i$. Then, $\Xi_i$ is compact and  invariant for \eqref{pf.s.xxi}.
 We now consider the interconnection between \eqref{pf.s.xxi} and the update laws \eqref{s.xi} for all $i\in\cN$, with the dynamics restricted to the invariant set $Z:=\Xi\x\R^N$, being $\Xi:=\prod_{i\in\cN} \Xi_i$. We compactly rewrite this interconnections as follows
 \begin{equation}\label{pf.s.z}
 z^{t+1} = \phi(z^t),\qquad z^t\in Z
 \end{equation}
 with $\phi$ suitably defined and  $z^t:=(\xi^t,x^t)\in\R^r\x\R^N$, being $\xi:=(\xi_i)_{i\in\cN}$ and $r:=\sum_{i\in\cN}\tau_i$.
Clearly, for every solution $x_a$ to \eqref{s.xi} starting at a given $t_0\in\N$ and subject to the signals $(h_i)_{i\in\cN}$, there is a   solution $z_b=(\xi_b,x_b)$ to \eqref{pf.s.z} starting at $0$ and such that $x_b(t)=x_a(t_0+t)$ for all $t\in\N$. 
For each compact $K\subset\Xi\x\R^N$, let $\cS(K)$ denote the set of solutions to \eqref{pf.s.z} starting at $0$ from $K$ and, for each $t\in\N$, define the reachable set from $K$ as  $\cR^t(K) := \big\{ (\xi^s,x^s)\in\Xi\x \R^{N}\st (\xi,x)\in\cS(K),\, s\ge t  \big\}$. In view of the above analysis, and since $\Xi$ is invariant for~\eqref{pf.s.z}, it follows that $\cR_{t}(K)$ is included in $\Xi\x\R^N$ and bounded uniformly in $K$ and $t$ for each $t\ge 1$. Thus,  the limit set $\Omega(K) := \bigcap_{t\in\N} \closure{\cR^t(K)}$ (where $\closure{\cR^t(K)}$ denotes the closure of $\cR^t(K)$) is compact, non-empty, and included in $\Xi\x\R^N$. Moreover,  since $\phi$ is continuous by construction, then $\Omega(K)$ is also forward invariant, uniformly globally attractive for \eqref{pf.s.z} from $K$ (see e.g. \cite[Proposition 6.26]{Goebel2012}), and it is the smallest set having the above properties. Furthermore, we notice that, by definition of the update laws \eqref{s.xi}, $x_i^t\in[\lb_{i},\M_i]$ for all $t\ge t_0$ despite the value of the initial conditions and of $t_0$, so that we conclude that $\Omega(K_1)=\Omega(K_2)$ for all $K_1,K_2$ supersets of $K\sr:=\prod_{i\in\cN} [\lb_{i},\M_i]$. In the following we let $\Omega:=\Omega(K\sr)$.

As $(h_i)_{i\in\cN}$ is uniformly exciting, by Claim 3 the convergence \eqref{ps.LS_xi2} holds  uniformly in the initial time. By the properties of $\Omega$, this implies that $\Omega\subset \Xi\x\cA_\epsilon$, and the projection $\cA_\epsilon^u  := \big\{ x\in\R^N\st (\xi,x)\in \Omega  \big \}$ 
satisfies $\cA_\epsilon^u\subset  \cA_\epsilon$. Therefore, it remains to show that $\cA_\epsilon^u$ is stable for $x$, i.e. that for each $\ell>0$, there exists $b(\ell)>0$, such that every solution to \eqref{pf.s.z} satisfying $\setdist{x^0}{\cA_\epsilon^u}\le b(\ell)$ also satisfies $\setdist{x^t}{\cA_\epsilon^u}\le \ell$ for all $t\in\N$. This, in turn, can be proved by similar arguments of \cite[Proposition 7.5]{Goebel2012}). In particular, suppose that the above stability property does not hold, and fix an $\ell>0$ arbitrarily. If $\cA_\epsilon^u$ is not stable, then for each $m\in\N$ there exist  $\tau_m\in\N$ and a solution $z_m=(\xi_m,x_m)\in\cS(Z)$  such that $\setdist{x^{0}_m}{\cA_\epsilon^u}\le 2^{-m}$ and $\setdist{x^{\tau_m}_m}{\cA_\epsilon^u}>\ell$. This, in turn implies
\begin{equation}\label{pf.Omg}
\setdist{z^{\tau_m}_m}{\Omega}>\ell.
\end{equation}  
Since $X_0:=\{x\in\R^N\st \setdist{x}{\cA_\epsilon^u}\le 1\}$ is compact, $Z_0:=\Xi\x X_0$ is compact. Thus, since $z^0_m\in Z_0$ for all $m\in\N$,  by uniform attractiveness of $\Omega$, there exists $\bar\tau=\bar\tau(\ell)\in\N$ such that $\tau_m\le \bar\tau$ for all $m\in\N$. We are thus given a sequence $(z_m|_{\le \bar\tau})_{m\in\N}$ of uniformly bounded  signals $z_m|_{\le \bar\tau}$, obtained by restricting the solutions $z_m$ to $\{0,\dots, \bar\tau\}$, which satisfies $\lim_{m\to\infty} \setdist{z^0_m}{\Omega} =0$. As $\phi$ is continuous, $Z$ is closed, and   since $\Omega$ is forward invariant, then in view of \cite[Theorem 6.8]{Goebel2012}  we can extract a subsequence of $(z_m|_{\le \bar\tau})_{m\in\N}$ (which we do not re-index)   that satisfies $\lim_{m\to\infty} \setdist{z^t_m}{\Omega} =0$ for all $t\in\{0,\dots,\bar\tau\}$. This, however, contradicts \eqref{pf.Omg} and proves the claim.

\subsection{Proof of Claim 5}
The last claim of the theorem, i.e. that if $(h_i)_{i\in\cN}$ is sufficiently exciting according to Definition \ref{d.SE} and $\lim_{t\to\infty} h_i^t=0$, then $\lim_{t\to\infty} x_i^t = \uM$ for all $i\in\cN$, follows directly from \eqref{pf.d.upsilon}-\eqref{pf.eq.lim_vep}.  \EOP

\appendix

\section{Proof of Lemma \ref{lem.PE}}\label{apd.lemma_PE}
For each $i\in\cN$, let $T_i\in\N_{\ge 1}$ be the period of $h_i$  and, with $t_i\sr$ and $h_i\sr>0$   such that $h_i^{t_i\sr}\ge  h_i\sr$, let
\[
r_i :=   t_i\sr  -   T_i\, \max\{ n\in\N\st T_i n \le   t_i\sr  \}.
\]
Then $r_i\in\{0,\dots, T_i\}$ and, since $h_i$ is $T_i$-periodic, for every $i\in\cN$ we have
\begin{equation}\label{pf.lemPE.hi}
h_i^{r_i + n T_i} \ge h_i\sr\qquad \forall n\in\N.
\end{equation}
Let $\Delta:= \max_{i\in \cN} T_i+1$ and $\und h:=\min_{i\in\cN} h_i\sr$. Fix arbitrarily   $m\in\N_{\ge 1}$ and $t_0\in\N$. Then we claim that, for each $i\in\cN$, there exists $n_i\in\N$ such that \[ s_i:= r_i + n_i T_i\in \Big\{t_0+1+ (m-1)\Delta,\,\dots,\,t_0+m\Delta  \Big\}.\]
In fact, if this is not true,   there  exist $m,t_0,n\in	\N$ and $i\in\cN$ such that 
  $r_i + n T_i < t_0+1+(m-1)\Delta$  and  $r_i +(n+1) T_i > t_0 + m\Delta$ hold. This, however, implies   
  \begin{align*}
  \Delta &= (1-m)\Delta + m\Delta <(1-m)\Delta + r_i +(n+1) T_i   - t_0 \\
  &< T_i +1,
  \end{align*}
which contradicts the fact that, by definition, $\Delta\ge T_i+1$ for all $i\in\cN$. Since \eqref{pf.lemPE.hi} implies that $h_i^{s_i}\ge \und h$ for all $i\in\cN$, then we claim that, for every $t_0\in\N$, $m\in\N_{\ge 1}$ (and thus, in particular, for those satisfying $m\le \log(\uM/\lb)/\und h$) and $i\in\cN$, there exists $s_i\in \{t_0+1+ (m-1)\Delta,\,\dots,\,t_0+m\Delta \}$ such that $h_i^{s_i}\ge \und h$, which proves the claim. \EOP{}

\section{Proof of Lemma \ref{lem.limsup}} \label{apd.Lemma_limsup}
	As $\nu\in[0,1)$, then for each $\epsilon  \in(0,1-\nu)$ there exists $t\sr_{1}\ge t_0$ such that
	\begin{align*}
	\nu^{t-t_0} x^{t_0} &\le \epsilon,&
	|y^t| &\le  \limsup_{t\to\infty} |y^t| + \epsilon,&
	\lambda^t & \le \limsup_{t\to\infty}\lambda^t + \epsilon 
	\end{align*}
	for all $t\ge t\sr_{1}$.
	As $\lambda^t\le \nu <1$ for all $t\ge t_0$, by iterating \eqref{e.lem.xy}, for $t>t\sr_{1}$, we obtain
	\begin{equation}\label{pf.lem.sum_x_1}
	\begin{aligned}
	&|x^t| \le \left(\prod_{s=t_0}^{t-1}\lambda^s \right) |x^{t_0}| + \sum_{s=t_0}^{t-1} \left(\prod_{\ell=s+1}^{t-1}\lambda^\ell \right) |y^s|\\
	&\le \nu^{t-t_0} |x^{t_0}| + \sum_{s=t_0}^{t\sr_{1}-1} \left(\prod_{\ell=s+1}^{t-1}\lambda^\ell \right) |y^s|  + \sum_{s=t\sr_1}^{t -1} \left(\prod_{\ell=s+1}^{t-1}\lambda^\ell \right) |y^s|\\
	&\le \epsilon + \sum_{s=t_0}^{t\sr_{1}-1} \nu^{t-s-1}   |y^s| + \sum_{s=t\sr_{1}}^{t -1} \left(\prod_{\ell=s+1}^{t-1}\lambda^\ell \right) |y^s|.
	\end{aligned} 
	\end{equation}
	As $y$ is bounded, there exists $c$ such that $|y^t|\le c$ for all $t\in\N$. Hence, the second term of the sum satisfies
	\begin{align*}
	\sum_{s=t_0}^{t\sr_{1}-1} \nu^{t-s-1}   y^s &= \nu^{t-t\sr_{1}}\sum_{s=t_0}^{t\sr_{1}-1} \nu^{t\sr_{1}-s-1}   y^s \le \nu^{t-t\sr_{1}}\dfrac{c}{1-\nu}.
	\end{align*}
	Therefore, there exists $t\sr_2\ge t\sr_1$ such that
	\begin{equation*}
	\sum_{s=t_0}^{t\sr_{1}-1} \nu^{t-s-1}   y^s \le \epsilon,\qquad\forall t\ge t\sr_2.
	\end{equation*}
	Denote for convenience $\bar y:=\limsup_{t\to\infty}|y^t|$ and $\bar\lambda:=\limsup_{t\to\infty}\lambda^t$.   As $\lambda^t\le \nu$ for all $t\ge t_0$, then $\bar \lambda\le\nu$. As $\epsilon<1-\nu$ by assumptions, then $\bar \lambda+\epsilon<1$. Therefore, since $t\ge t\sr_1$, then the last term of \eqref{pf.lem.sum_x_1} satisfies
	\begin{equation}
	\begin{aligned}
	&\sum_{s=t\sr_{1}}^{t -1} \left(\prod_{\ell=s+1}^{t-1}\lambda^\ell \right) |y^s| \le \sum_{s=t\sr_{1}}^{t -1} \big(\bar \lambda +\epsilon \big)^{t-s-1} \big(\bar y+\epsilon\big)\\
	&\le \dfrac{\epsilon}{1-\nu} + \dfrac{\bar y}{1-(\bar \lambda+\epsilon)}\\
	&\le \dfrac{\bar y}{1- \bar \lambda } + \dfrac{\epsilon}{1-\nu} + \dfrac{\bar y}{1-(\bar \lambda+\epsilon)}-\dfrac{\bar y}{1- \bar \lambda }\\
	&\le \dfrac{\bar y}{1- \bar \lambda } +    \bar y p(\epsilon)
	\end{aligned}
	\end{equation}
	in which $p:[0,1-\nu)\to\R$,  defined as
	\[
	p(\epsilon) := \dfrac{\epsilon}{1-\nu} + \dfrac{\epsilon}{(1-\bar\lambda)(1-\bar\lambda-\epsilon)},
	\]
	  is continuous and satisfies $\lim_{\epsilon\to 0} p(\epsilon) = 0$.
	From \eqref{pf.lem.sum_x_1} we get $
	|x^t|\le  \bar y/(1- \bar \lambda ) +  2 \epsilon + \bar y p(\epsilon)$ 
	for all $t\ge t\sr_{2}$, and the claim follows by arbitrariness of $\epsilon$. 
	\EOP{}

  
  \bibliography{biblio}

  \addtolength{\textheight}{-12cm}   

\end{document}